\documentclass[preprint, superscriptaddress, preprintnumbers,amsmath,amssymb]{revtex4}

\usepackage{graphicx}
\usepackage{dcolumn}
\usepackage{bm}
\usepackage{textcomp}
\usepackage{xcolor}
\usepackage[normalem]{ulem}
\usepackage{soul}
\usepackage{subfigure}
\usepackage{array}
\usepackage{footnote}
\usepackage{makecell}
\usepackage{verbatim}
\usepackage{threeparttable}
\usepackage{float}
\usepackage{appendix}
\usepackage{amsmath,amssymb}
\usepackage{cancel}
\usepackage{tipa}
\usepackage{flushend}
\usepackage{booktabs}
\usepackage{color,graphicx}
\usepackage{amsmath}
\usepackage{dcolumn}
\usepackage{CJK}               
\usepackage{bm}            
\usepackage{soul}
\usepackage{enumerate}
\usepackage{multirow}
\usepackage{mathrsfs}
\usepackage{braket}
\usepackage{longtable}
\usepackage{pdflscape}
\usepackage[pdfstartview=FitH,
            CJKbookmarks=true,
            bookmarksnumbered=true,
            bookmarksopen=true,
            colorlinks,
            pdfborder=001,
            linkcolor=blue,
            anchorcolor=blue,
            citecolor=blue
            ,pdftex]{hyperref}
\usepackage{float}

\begin{document}
%\begin{CJK*}{GBK}{song}
\title{Point-proton density distributions of stable nuclei}
\author{Tian Yu Wu}
 \affiliation{School of Physics, Beihang University, Beijing 100191, China}
\author{Bao Hua Sun}
\email{bhsun@buaa.edu.cn}
\affiliation{School of Physics, Beihang University,  Beijing 100191, China}
\author{Hui Hui Xie}
\affiliation{College of Physics, Jilin University, Changchun 130012, China}
 \author{Jun Yao Xu}
 \affiliation{School of Physics, Beihang University, Beijing 100191, China}
 \author{Ge Guo}
 \affiliation{School of Physics, Beihang University, Beijing 100191, China}
\date{\today}

\begin{abstract}
  Point-proton density distributions are deduced for 130 stable nuclei from $^{7}\mathrm{Li}$ to $^{232}\mathrm{Th}$ from nuclear charge densities determined in elastic electron scattering. There are 171 cases presented in model-dependent forms, including the modified Harmonic-oscillator function, two-parameter Fermi function (2pF), three-parameter Fermi function, three-parameter Gaussian function, and 97 in Fourier-Bessel series model-independent forms. Independent of density functions, the point-proton root-mean-square (rms) radii of the derived point-proton density show excellent agreement with each other. We identify cases where the tabulated data of charge densities and charge radii are inconsistent, and the deduced point-proton density distributions are inaccurate due to insufficient experimental momentum transfer coverage or inconsistent scattering experiments. For the widely used 2pF distribution, it is found that the surface diffuseness parameters can be empirically calculated from those of charge density, while the half-density radius parameters follow the $A^{1/3}$ rule. The derived point-proton density distributions can be used as input in nuclear reaction studies and compared with nuclear model predictions. 
\end{abstract}

\date{\today}

\maketitle
\tableofcontents

\section{Introduction} \label{section1}
In the study of nuclei---finite quantum systems with diffuse boundaries---the point-proton density distribution ($\rho_p$) plays a fundamental role. It not only captures the overall size and detailed internal distribution of protons but also serves as a cornerstone for defining key physical quantities, such as point-proton root-mean-square (rms) radii ($\langle r^2\rangle_p^{1/2}$). In addition, it is also an indispensable input for nuclear reaction models to link nuclear structure with reaction mechanisms. A systematic study of the nucleon density distributions of finite nuclei is important for understanding the isospin dependence of the nuclear many-body system. 

Experimentally, point-proton density distributions have been extracted through proton elastic scattering~\cite{SAKAGUCHI20171, Terashima, Zenihiro2010PRC} and heavy ions reactions~\cite{jichao2024, jiwneiPLB2024, KaurPRL}. However, utilizing the hadronic probes is challenging due to the interplay between the nuclear structure and the reaction dynamics governed by the strong force. 
On the other hand, electromagnetic probes that are well-understood by quantum electrodynamics (QED) 
have been employed for measuring the charge density distribution ($\rho_c$). On the order of the fine-structure constant $\alpha$=1/137, the weakly interacting nature between the electron and target nuclei prevents disruption of the target nuclei. Being a useful technique for studying nuclear structure~\cite{walecka_2023,sudaPPNP, Liu_2019}, elastic electron scattering remains the sole method for ascertaining the charge density distribution $\rho_c$ within a nucleus. Until now, elastic electron scattering experiments have targeted almost all the stable nuclei~\cite{DEVRIES1987495, FRICKE1995177}. Recently, the electron scattering has been successfully extended to unstable nuclei~\cite{PhysRevLett.131.092502}. 

Nuclear charge density distribution is primarily determined by the distribution of point protons inside an atomic nuclide. This work aims to provide a consistent database of point-proton density distributions for stable isotopes from elastic electron scattering experiments. This database can serve as a basic input for nuclear reaction models and examinations for nuclear structure models. For this purpose, the point-proton densities are derived in different forms neglecting the contribution of spin-orbital density. The point-neutron density ($\rho_n$) is assumed to be $\rho_n=\frac{N}{Z}\rho_p$. 

This paper is organized as follows. In Sec .~\ref {sec2}, we will survey the charge density data, compare different density distribution forms, and present the details of the $\rho_p$ deduction procedure. In Sec.~\ref{sec3}, we discuss the deduced $\rho_p$ compared to $\rho_c$, the precisions in point-proton rms radii and $\rho_p$ for $^{48}\mathrm{Ca}$ and $^{208}\mathrm{Pb}$. Finally, the conclusion is given in Sec .~\ref {sec4}.

%%%%%%%%%%%%%%%%%%%%%%%%%%%%%%%%%%%%%%%%%%%%%%%%%%%%%%%%%%
%                    begin  theoretical framework
%%%%%%%%%%%%%%%%%%%%%%%%%%%%%%%%%%%%%%%%%%%%%%%%%%%%%%%%%%

%----------------------------------------------------------------------------------------
\section{Derivation of the Point-proton density distribution}
\label{sec2}

\subsection{Overview of nuclear charge density  data}\label{subsec1}

In elastic electron scatterings, the differential cross sections $\frac{d\sigma}{d\Omega}$ are connected to the nuclear response, the charge form factor ($F_{c}$). It is the Fourier component of the charge density distribution at a specific momentum transfer, $q$, in the framework of plane-wave impulse approximation (PWIA)~\cite{sudaPPNP, Suda:2023msg}. Owing to the dependence of the Mott cross section on $1/q^4$, the determination of $F_{c}$, in reality, has to be measured in a limited range of momentum transfer. This necessitates employing a phenomenological model to parameterize the charge density distribution effectively. The ``experimental" charge densities~~\cite{DEVRIES1987495, FRICKE1995177} have been tabulated in model-dependent forms, including the Harmonic-oscillator function (HO), Modified Harmonic-oscillator function (MHO), two-parameter Fermi function (2pF), three-parameter Fermi function (3pF), three-parameter Gaussian function (3pG), model-independent analyses of Sum-of-Gaussian function (SOG)~\cite{SICK1974509} and Fourier-Bessel (FB)~\cite{Dreher}. The explicit expressions are given in the Appendix~\ref{app1}.

% In most cases, the Harmonic-oscillator(HO) density distribution provides a good approximation for light nuclei~\cite{bertulani2001physics}, which is one of the most widely used distributions for describing both charge and nucleon density distribution of light nuclei in p-shell~\cite{DEVRIES1987495, KaurPRL, TranPRC}. For medium to heavy nuclei, the fermi and Gaussian distribution are commonly employed with two or three free parameters for both stable and unstable nuclei~\cite{TanakaPRL, KanungoPRCMg, NPA2020, NeutronrearrangementPLB2023}, which are subsequently referred to as the two-parameter fermi(2PF), three-parameter fermi(3PF) and three-parameter Gaussion(3PG) functions in the latter of the paper. Compared to the previous density distribution forms, the density distribution expanded into the Fourier-Bessel series and the sum of Gaussian enable model-independent analyses. The comparison of these distributions is presented in~Sec. \ref{unfolding}. 

% \textcolor{red}{Since the two parameters in 2PF are separately responsible for the dip position and the height of the diffraction maximum of the form factor, }
\begin{figure}[h]
\centering
\includegraphics[width=0.5 \textwidth]{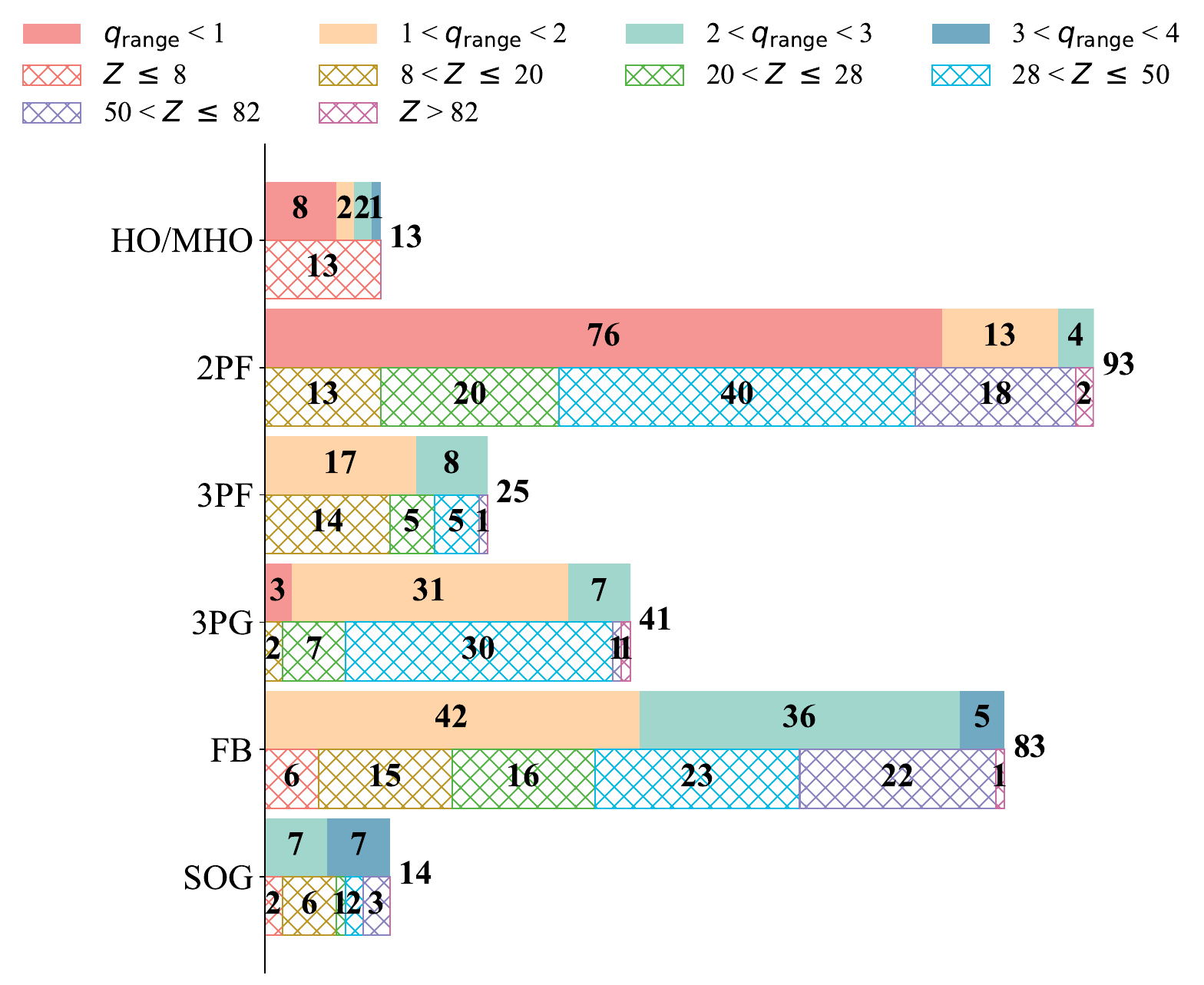}
\caption{\label{fig:range_disrtibutions}
Classification of charge density data in Refs.~\cite{DEVRIES1987495, FRICKE1995177} by the density distribution models, the momentum transfer ranges covered and the proton numbers ($Z$).
%specific shell nuclei used in each model. 
Shown in each category is the number of cases. 
Note that there are cases where various experiments exist for the same nucleus.}
\end{figure} 

Charge density distribution parameters and charge radii were compiled for 136 stable nuclei from $^{3}\mathrm{H}$ to $^{238}\mathrm{U}$ in Refs.~\cite{DEVRIES1987495, FRICKE1995177}.  All the $\rho_c$ data are categorized in Fig.~\ref{fig:range_disrtibutions} according to the momentum transfer range ($q_\text{range}$) covered experimentally and the proton numbers of relevant nuclides. A significant portion of the charge density data expressed in HO/MHO and 2pF forms corresponds to $q_\text{range}$ of less than 1 $\textrm{fm}^{-1}$. In this case, the form factor covering the first maximum can help to evaluate the gross size and shape of $\rho_c$. For 3pF and 3pG forms, the additional inner depth parameter offers more details of the density distributions. The form factors often reach the second minimum, with most $q_\text{range}$ lying between 1 and 2 $\textrm{fm}^{-1}$. 
For the model-independent distributions, a larger $q_\text{range}$ is typically required to look into the details of the central region of the density. As elucidated in Fig.~\ref{fig:range_disrtibutions}, nearly half of the FB data, namely 41 out of 83, are measured with $q_\text{range}$ between 2 and 4 $\textrm{fm}^{-1}$, while the remaining data have $q_\text{range}$ less than 2 $\textrm{fm}^{-1}$. For the latter, the $q^{-4}$exp$(-q^{2}r^2_p)$ dependence is assumed for $F_c$ beyond the experiment accessible data~\cite{DEVRIES1987495, Dreher},  where $r^2_p$ is the mean-square charge radii of a proton.

%If classified by different shells, 
For the $p$-shell nuclei, the simple model-dependent density forms, HO and MHO, are usually employed as an effective approximation.
%for light nuclei with proton numbers less than 8~\cite{bertulani2001physics}. 
The most adapted 2pF form has been used to describe nuclei heavier than oxygen, together with 3pF and 3pG.  %Fermi (rank the first in terms of quantity) and Gaussian distribution extend to Z $>$ 8 up to Z = 82. 
To avoid model dependency, model-independent distributions, SOG, and FB are introduced. In particular, 83 cases in FB form are supposed to be ``ideal" charge density data. 
%Both the model-independent SOG and FB distributions (most prevalently implemented to reduce model dependency) are employed to depict density distribution for both light and heavy nuclei. 
It is worth noting that different experiments exist for one nucleus. 
% \textcolor{red}{You should comment: do these experiments agree with each other or not! For example, these experiments often differ in the labs, where the $ q$ range is covered and the density form employed. However, they tend to give similar charge radii.}.
%We follow the same instructions for proton density distribution. 
% In total, over 200 charge densities,  13 in HO/MHO, 93 in 2PF, 25 in 3PF, 40 in 3PG, 83 in FB, and 14 in SOG are utilized in this work to deduce the point-proton density distributions for xx isotopes.
We noticed some data exhibit inconsistencies, as the rms charge radii $\langle r^2\rangle_\mathrm c^{1/2}$ tabulated in Refs.~\cite{DEVRIES1987495, FRICKE1995177} deviate from those calculated with the tabulated density parameters. The data with radii deviations over 0.01 $\mathrm{fm}$ are listed in Table~\ref {tab:wrongdata}.

\begin{table}[h]
\caption{\label{tab:wrongdata}
Inconsistent data in Refs.~\cite{DEVRIES1987495,FRICKE1995177}. Shown
are the specific nuclide, charge density distribution form, the tabulated rms charge radii
 ($^{\mathrm{table}}\langle r^2\rangle_\mathrm c^{1/2}$), the calculated rms charge radii ($^{\mathrm{cal}}\langle r^2\rangle_\mathrm c^{1/2}$) using the tabulated charge densities and the deviation of the above two.}
 \renewcommand{\arraystretch}{0.8}
\begin{ruledtabular}
\begin{tabular}{ccccc}
\textrm{Nucleus}&
\textrm{Model}&
\parbox{2cm}{$^{\mathrm{table}}\langle r^2\rangle_\mathrm c^{1/2}$ \\ (fm)}&
\parbox{2cm}{$^{\mathrm{cal}}\langle r^2\rangle_\mathrm c^{1/2}$ \\ (fm)}&
% \textrm{$^{table}\langle r^2\rangle_\mathrm c^{1/2}$ (fm)}&
% \textrm{$^{cal}\langle r^2\rangle_\mathrm c^{1/2}$ (fm)}&
\parbox{2cm}{Deviation \\ (fm)}\\
\hline
$^{20}\mathrm{Ne}$&3pF&2.992\footnotemark[1]&2.949&0.043\\
$^{24}\mathrm{Mg}$&3pF&2.985\footnotemark[1]&2.956&0.029\\
$^{25}\mathrm{Mg}$&3pF&3.003\footnotemark[1]&2.982&0.021\\
$^{28}\mathrm{Si}$&3pF&3.086\footnotemark[1]&3.076&0.01\\
$^{48}\mathrm{Ti}$&2pF&3.713\footnotemark[1]&3.693&0.02\\
$^{59}\mathrm{Co}$&3pG&3.775\footnotemark[1]&3.629&0.146\\
$^{89}\mathrm{Y}$&2pF&4.24\footnotemark[1]&4.254&0.014\\
$^{96}\mathrm{Zr}$&3pG&4.396\footnotemark[1]&4.371&0.025\\
$^{93}\mathrm{Nb}$&2pF&4.31\footnotemark[1]&4.332&0.022\\
$^{74}\mathrm{Se}$&2pF&4.07\footnotemark[2]&4.081&0.011\\
$^{74}\mathrm{Se}$&3pG&4.04\footnotemark[2]&4.131&0.091\\
$^{76}\mathrm{Se}$&3pG&4.133\footnotemark[2]&4.212&0.079\\
$^{78}\mathrm{Se}$&3pG&4.111\footnotemark[2]&4.196&0.085\\
$^{82}\mathrm{Se}$&3pG&4.122\footnotemark[2]&4.18&0.058\\
\end{tabular}
\end{ruledtabular}
\footnotetext[1]{From dataset~\cite{DEVRIES1987495}.}
\footnotetext[2]{From dataset~\cite{FRICKE1995177}.}
\end{table}

\subsection{Charge density distributions in different forms}\label{subsec2}

\begin{figure}[!htpb]
\centering
\includegraphics[width=0.45\textwidth]{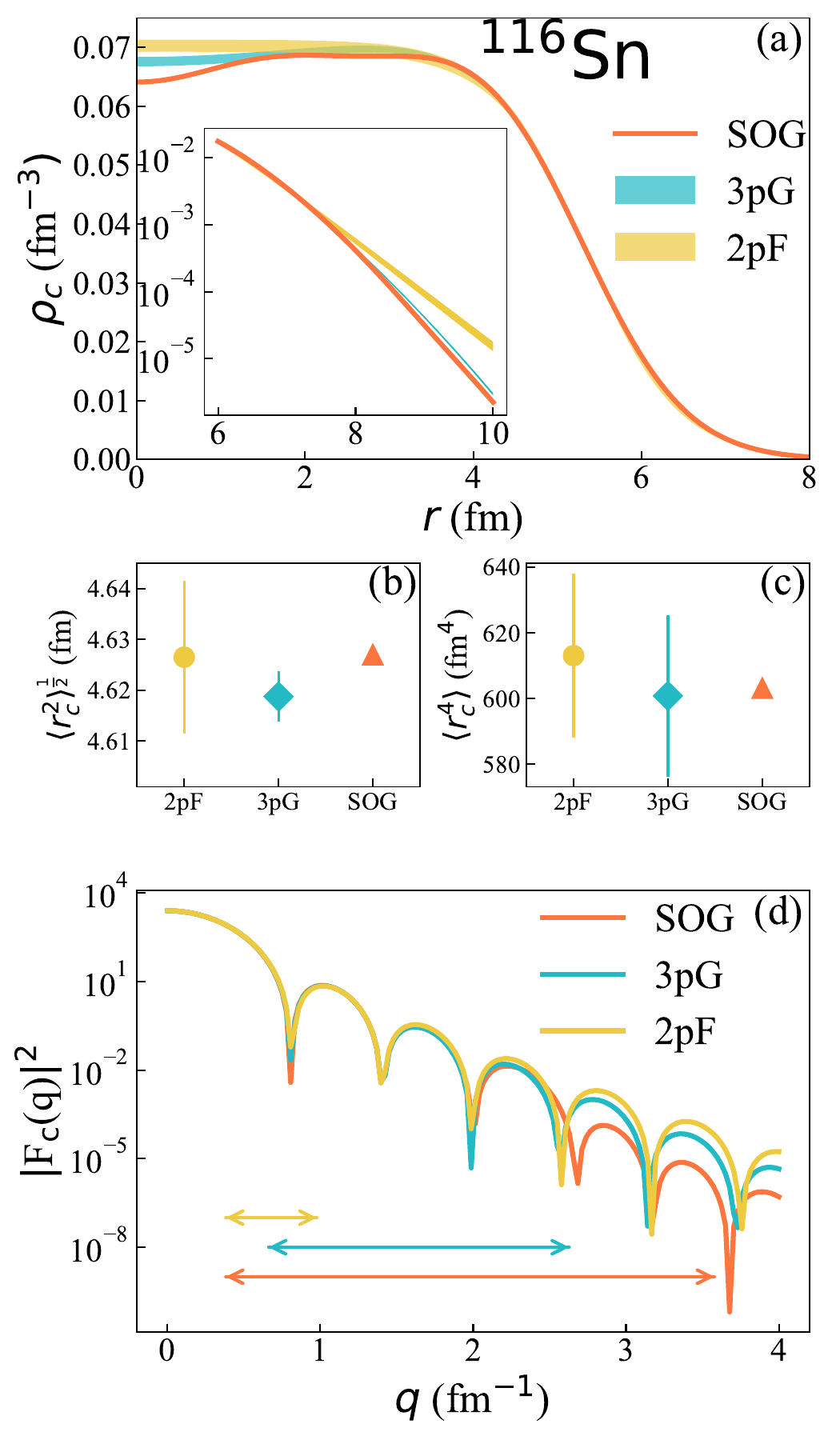}
\caption{\label{fig:plot_range}
(a) Comparison of $\rho_c$ of $^{116}$Sn measured in different experiments in Ref.~\cite{DEVRIES1987495} with three function types: 2pF (blue line), 3pG (orange line), and SOG (red line). The inset shows the densities at the tail region. The relevant rms radii (b) and fourth-order moment (c) of three densities are also given. (d) The corresponding form factor $F_c$ is deduced from $\rho_c$. Arrows represent the momentum transfer regions covered in the experiments.}
\end{figure}

For 75 cases out of 136 nuclei in the data compilations, different charge densities have been extracted from different experiments for one single nuclide. One such case is $^{116}\mathrm{Sn}$.     %\textcolor{brown}{Multiple experimental measurements are available for 75 out of 136 nuclei, often covering different momentum transfer ranges and providing various forms of density distributions.} 
%Thus we first examine the specific cases where different charge densities are obtained from different experiments for one single nuclide like $^{116}$Sn. 
In Fig.~\ref{fig:plot_range} (a), the charge densities of $^{116}\mathrm{Sn}$ are presented in three different forms, i.e., the model-independent SOG~\cite{116SnSOG} with 12 free parameters, the model-dependent 2pF~\cite{116Sn2PF} and 3pG~\cite{116Sn3PG} models. The error bands for 2pF and 3pF are also indicated. For SOG, no uncertainties are given in the original literature. 

In contrast to the uniform distribution of the 2pF model, the SOG displays a central density resembling a “wine bottle”. The differences among the three densities start to reduce at the shoulder region and become hardly discernible at the tail region. 
%The 3PG follows well the decreasing trend of SOG towards the density tail, but 2PF starts to deviate from SOG at around $r > 8 fm$.
Although 2pF and 3pG cannot reproduce the details in the central area, they can give consistent $\langle r^2 \rangle_{c}^{1/2}$ and $\langle r^4\rangle_{c}$ values within experiment uncertainties as seen in Fig.~\ref{fig:plot_range} (b) and (c). 
% Recently, the fourth-order moment of the charge distribution, $\langle r^2 \rangle_{c}^{4}$, has attracted much attention due to its potential in extracting neutron rms radii~\cite{PhysRevC.101.021301, nthorder, XIE2023PLB}. As shown in Fig.~\ref{fig:plot_range} (c), all three forms give consistent $\langle r^4\rangle_{c}$ values within error bars. 

In Fig.~\ref{fig:plot_range} (d), the charge form factors $F_c$ are presented for the three types of densities. High-momentum transfer scattering explores the internal density distribution of the nucleus, whereas low-momentum transfer focuses on the surface area. In contrast to the SOG model, the 2pF and 3pG form factors begin to deviate significantly when $q >$ 2.5 fm$^{-1}$, highlighting their limitations in representing the central density.
% 2pF and 3pG form factors deviate from SOG when $q >$ 2.5 fm$^{-1}$, reflecting their insufficiency to describe the central density. 
The least range of $q$ would ensure 2pF and 3pG to characterize the nucleus bulk and shape, %~\cite{sudaPPNP}, 
but it is risky to extend 2pF and 3pG form factors to the $q_\text{range}$ beyond experiments. 
Therefore, one should note that $\rho_p$ should be deduced in the same $q_\text{range}$ as the experiment covered.  

\subsection{Unfolding charge density distributions}\label{subsec:expression}

The nuclear charge density distribution in the momentum space $F_c$, is written as~\cite{xie2023impact}
\begin{equation}\label{fourier}
    F_{\rm c}(\bm q)=\sum_{\tau\in\{p,n\}}\left[G_{\mathrm E\tau}(\textbf{\textit{q}}^2)F_{\tau}(\bm q)+F_{2\tau}(\textbf{\textit{q}}^2)F_{W\tau}(\bm q)\right] \;,
\end{equation}
where $F_\tau$ and $F_{W\tau}$ ($\tau\in \{p,n\}$) are the Fourier transform of point-nucleon density and the spin-orbit density~\cite{xiePRC2024,xiePRA},  respectively. $\textbf{\textit{q}}^2 \equiv - q_{\mu}q^{\mu}$, where $q_\mu$ is the momentum transfer. 
The quantities $G_{E\tau}$ and $F_{2\tau}$ are the nucleon's electric Sachs and Pauli form factors, respectively.
% $G_{E\tau}$ and $F_{2\tau}$ are the electric Sachs and Pauli form factors of a nucleon, respectively.

The parameterization of nucleon form factors adopted in this work are taken from Ref.~\cite{formfactorPRC}, 
\begin{equation}\label{formfactor}
\begin{aligned}
    G_{Ep}(q^2) &= \frac{1}{(1+r_{p}^2\mathbf{\textit{q}}^2/12)^2} \;,\\
    G_{En}(q^2) &= \frac{1}{(1+r_{+}^2\mathbf{\textit{q}}^2/12)^2} - \frac{1}{(1+r_{-}^2\mathbf{\textit{q}}^2/12)^2} \;,
\end{aligned}
\end{equation}
where the proton charge radius $r_p$ = 0.8414 fm~\cite{protonradius} and $r^2_{\pm}=r_{av}^2$ $\pm$ $\frac{1}{2} r^2_{n}$, where $r^2_{av}$ = 0.81 fm$^2$ is the average of the squared radius for positive and negative charge distributions and $r_n^2 =-0.11$~fm$^2$~\cite{neurad} is the mean-square radius of the neutron.

The nuclear charge density can be derived from the inverse Fourier transformation of the nuclear charge form factor, %~\eqref{chargeexpress}. 
%Thus, the charge density is expressed by considering the contributions from both point-proton and -neutron density, the proton and neutron spin-orbit densities, and the single-proton and single-neutron charge densities. 
and is written in the relativistic form as:
\begin{equation}
    \label{chargeexpress}
    \rho_{\rm c}(r)=\sum_{\tau\in\{p,n\}}\left[\rho_{\mathrm c\tau}(r)+W_{\mathrm c\tau}(r)\right],
\end{equation}
where
\begin{align}\label{eq-rct}
    &\rho_{\mathrm c\tau}(r)=\frac1r\int_0^\infty\mathrm dx x\rho_\tau(x)\left[g_\tau(|r-x|)-g_\tau(r+x)\right],\\
    &W_{\mathrm c\tau}(r)=\frac1r\int_0^\infty\mathrm dx xW_\tau(x)\left[f_{2\tau}(|r-x|)-f_{2\tau}(r+x)\right].
\end{align}
Here, $\rho_\tau(x)$ and $W_\tau(x)$ are the nucleon and spin-orbit density. The functions $g_\tau(x)$ and $f_{2\tau}(x)$ are the Fourier transforms of Sachs and Pauli form factors, respectively. The charge density depends on the point-proton and point-neutron density, the proton and neutron spin-orbit densities, and the single-proton and single-neutron charge densities. 

The corresponding mean-square radius of nuclear charge density for the nuclide ($Z$, $N$),  $\langle r^2 \rangle_{c}$, is given by~\cite{PhysRevC.86.045503}
\begin{equation}\label{rch}
    \langle r^2\rangle_c=\langle r^2\rangle_p+r^2_p+\langle r^2\rangle_{\mathrm{W_{p}}}+\frac NZ\left(r_n^2+\langle r^2\rangle_{\mathrm{W_{ n}}}\right) \;,
\end{equation}
where $Z$ and $N$ are the number of protons and neutrons, respectively. The term $\langle r^2\rangle_p$ is the point-proton mean-square radius. The terms $\langle r^2\rangle_{\mathrm{W_{p}}}$ and $N/Z\langle r^2\rangle_{\mathrm{W_{n}}}$ represent the proton and neutron spin-orbit contributions, respectively. Due to the proton charge radius puzzle~\cite{Gao2024}, adopting the alternative value of $r_p = 0.88$ fm would lead to a difference in $\langle r^2\rangle_{p}^{1/2}$ of approximately $0.0688 / (2\langle r^2\rangle_c^{1/2}) \approx 0.0688 / (2.4 A^{1/3})$ fm, $i.e.$, 0.015 fm for $A$ $\leq$ 10, and 0.006 fm for $A$ $\approx$ 200. The differences are generally smaller or comparable to experimental uncertainties. 
The deduced $\rho_p$ is insensitive to the selection of $r_p$.
% Neglecting  the spin-orbit densities in Eq.~\eqref{fourier} results in $\langle r^2\rangle_{\mathrm{W_{\rm p}}}$ and $N/Z\langle r^2\rangle_{\mathrm{W_{\rm n}}}$ vanish. 

% The Eq.\eqref{chargeexpress} is then calculated using a Fourier transform as:
% \begin{equation}
%     F_{\rm c}(\bm q)=\int\mathrm d^3\bm r e^{i\bm q\cdot\bm r}\sum_{\tau\in\{p,n\}}\left[G_{\mathrm E\tau}(Q^2)\rho_\tau( r)+F_{2\tau}(Q^2)W_\tau( r)\right].
% \end{equation}

\begin{figure}[!htpb]
\centering
\includegraphics[width=0.32\textwidth]{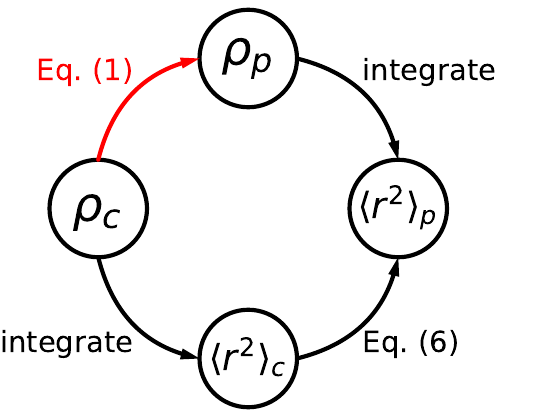}
\caption{\label{fig: flowchart}
Relationship of charge density distribution ($\rho_{c}$), point-proton density distribution ($\rho_{p}$), mean-square charge radii ($\langle r^2 \rangle_{c}$) and mean-square point-proton radii ($\langle r^2 \rangle_{p}$).
}
\end{figure}

In Fig.~\ref{fig: flowchart}, $\rho_{c}$, $\rho_{p}$, $\langle r^2 \rangle_{c}$ and $\langle r^2 \rangle_{p}$ are illustrated with equations interconnected. This work aims to provide systematic point-proton density distributions deduced from $\rho_c$, as indicated by the red arrow. The unique $\langle r^2 \rangle_p$ value can be calculated from Eq.~\eqref{rch} or from integrating the deduced $\rho_{p}$. 
%\textcolor{brown}{Note that they are strictly identical~\cite{nthorder}.}. 
Naturally, the deviations in $\langle r^2 \rangle_p$ between the two ways point to the accuracy of computing $\rho_c$ to $\rho_p$. 

% The deviation between $\langle r^2 \rangle_p$ derived from the predefined distribution function and that from Eq.\eqref{rch} reflects the accuracy of the deduced $\rho_p$. Notably, Eq.\eqref{rch} is model-independent, making it a reliable reference for comparison.
% Thus the deviation between $\langle r^2 \rangle_p$ from deduced $\rho_p$ with a predefined distribution function and from Eq.~\eqref{rch} indicate the rms radii accuracy of the deduced $\rho_p$ since no model assumption is made when deducing Eq.~\eqref{rch}. More details can be found in Ref.~\cite{nthorder}.

Note that the original $F_c$ data is often missing from the literature. Instead, the tabulated $\rho_c$ is employed to deduce $F_c$ in this work. The point-proton densities are deduced from the corresponding charge densities, which are parameterized in the same form. For $\rho_c$ in the FB and SOG forms, we deduce $\rho_p$ in the FB form. For a systematic study, we provide $\rho_p$ in the 2pF form for all the $\rho_c$ data in the 2pF, 3pF, 3pG, FB, and SOG forms.
% $\rho_p$ are deduced from the relevant $\rho_c$ parameterized in the same form for each data. For example, for cases where $\rho_c$ is 3pF, $\rho_p$ is also calculated in 3pF form.  
% $\rho_p$ in FB are employed for $\rho_c$ in FB and SOG. 
% For a systematic study, 2pF $\rho_p$ are provided for all the $\rho_c$ data in 2pF, 3pF, 3pG, FB, and SOG forms. 

The free parameters in $\rho_p$ are determined by fitting $F_c$ with the nonlinear least-squares method~\cite{SciPy-NMeth}. The $q_{\mathrm{range}}$ is fixed in the same way as covered in electron scattering. The point-neutron density $\rho_n$ is assumed to be $\frac{N}{Z}\rho_p$~\cite{Brown2014}. We have examined the impact of the neutron skin effect by taking $^{48}\mathrm{Ca}$, $^{116}\mathrm{Sn}$, and $^{208}\mathrm{Pb}$ as examples, which have known $\langle r^2 \rangle_{n}^{1/2}$~\cite{zenihiro2018, Terashima, Zenihiro2010PRC}. We found it has minimal effect on $\langle r^2 \rangle_{p}^{1/2}$ and $\rho_p$. % with $\rho_c$ in the FB, SOG, and FB forms, respectively. These nuclei have $\langle r^2 \rangle_{n}^{1/2}$ determined from proton elastic scattering~\cite{zenihiro2018, Terashima, Zenihiro2010PRC}. The neutron density distributions are assumed in 2pF form, with radius parameters~\eqref{2pf} adjusted to reproduce the measured $\langle r^2 \rangle_{n}^{1/2}$. The surface diffuseness parameters are equal to those of $\rho_p$. As a result, the final $\langle r^2 \rangle_{p}^{1/2}$ increases by only 0.0003 fm, 0.0004 fm, and 0.0026 fm for $^{48}\mathrm{Ca}$, $^{116}\mathrm{Sn}$, and $^{208}\mathrm{Pb}$, respectively, which are below or at their experimental uncertainties of 0.009 fm, 0.001 fm and 0.002 fm.}
% The changes for $^{48}\mathrm{Ca}$ and $^{116}\mathrm{Sn}$ are smaller than their respective experimental uncertainties (0.009 fm and 0.001 fm), while the change for $^{208}\mathrm{Pb}$ slightly exceeds its experimental uncertainty (0.002 fm).} 
One should note that the spin-orbit densities of neutrons and protons have opposite signs, thus partially offset each other in the overall contribution~\cite{xie2023impact}. For the above reason, the contribution of spin-orbit densities is neglected, so that $\langle r^2\rangle_{\mathrm{W_{\rm p}}}$ and $N/Z\langle r^2\rangle_{\mathrm{W_{\rm n}}}$ are thrown out of Eq.~\eqref{rch}.

% \textcolor{blue}{For the existing proton charge radius puzzle~\cite{Gao2022, Gao2024}, we also test $^{48}\mathrm{Ca}$ and $^{208}\mathrm{Pb}$ by changing $r_p$ to 0.88 fm in Eq.~\eqref{formfactor}. The resulting $\langle r^2 \rangle_{p}^{1/2}$ decrease 0.017 and 0.013 fm, respectively. These variations are slightly larger than the experimental uncertainties of 0.009 fm, 0.009 fm, 0.002 fm, indicating that the final results is somewhat sensitive to the choice of $r_p$. }
\iffalse
\textcolor{brown}{The unknown parameters of point-proton density distributions are determined by fitting the calculated charge form factor $F_c$ obtained from the experimental $rho_c$ with a method of nonlinear least-squares fit~\cite{SciPy-NMeth} as the same range as the electron scattering covered. The neutron density in Eq.~\eqref{fourier} is to be $\rho_n=\frac{N}{Z}\rho_p$, which is widely used, e.g., in the proton scattering experiments. The spin-orbit densities are also neglected in Eq.~\eqref{fourier}, resulting in a null $\langle r^2\rangle_{\mathrm{W_{\rm p}}}$ and $N/Z\langle r^2\rangle_{\mathrm{W_{\rm n}}}$ in Eq.~\eqref{rch}. See Ref.~\cite{nthorder} for more details. One should note that the spin-orbit densities of neutrons and protons have opposite signs and will cancel each other out, resulting in a minimal overall contribution~\cite{xie2023impact}. } 
\fi

Point-proton density distributions are deduced from over 200 charge densities for 136 stable nuclei~\cite{DEVRIES1987495, FRICKE1995177}. Among them, 13 $\rho_c$ cases are parameterized in HO/MHO, 93 in 2pF, 25 in 3pF, 40 in 3pG, 83 in FB, and 14 in SOG. 
% \textcolor{red}{The two lightest nuclei ($^{3}\mathrm{H}$ and $^{3}\mathrm{He}$), $^{23}\mathrm{Na}$ with Uniform Gaussian distribution and three deformed nuclei ($^{166}\mathrm{Er}$, $^{176}\mathrm{Yb}$ and $^{238}\mathrm{U}$) are excluded in this work}.
All the derived $\rho_p$ parameters and radii are listed in Tables~\ref {tab:grossdata} and~\ref{tab:grossdatafb}.

\section{results and discussion}\label{sec3}

\subsection{Accuracy of point proton rms radii}

\begin{figure}[h]
\centering
\includegraphics[width=0.5\textwidth]{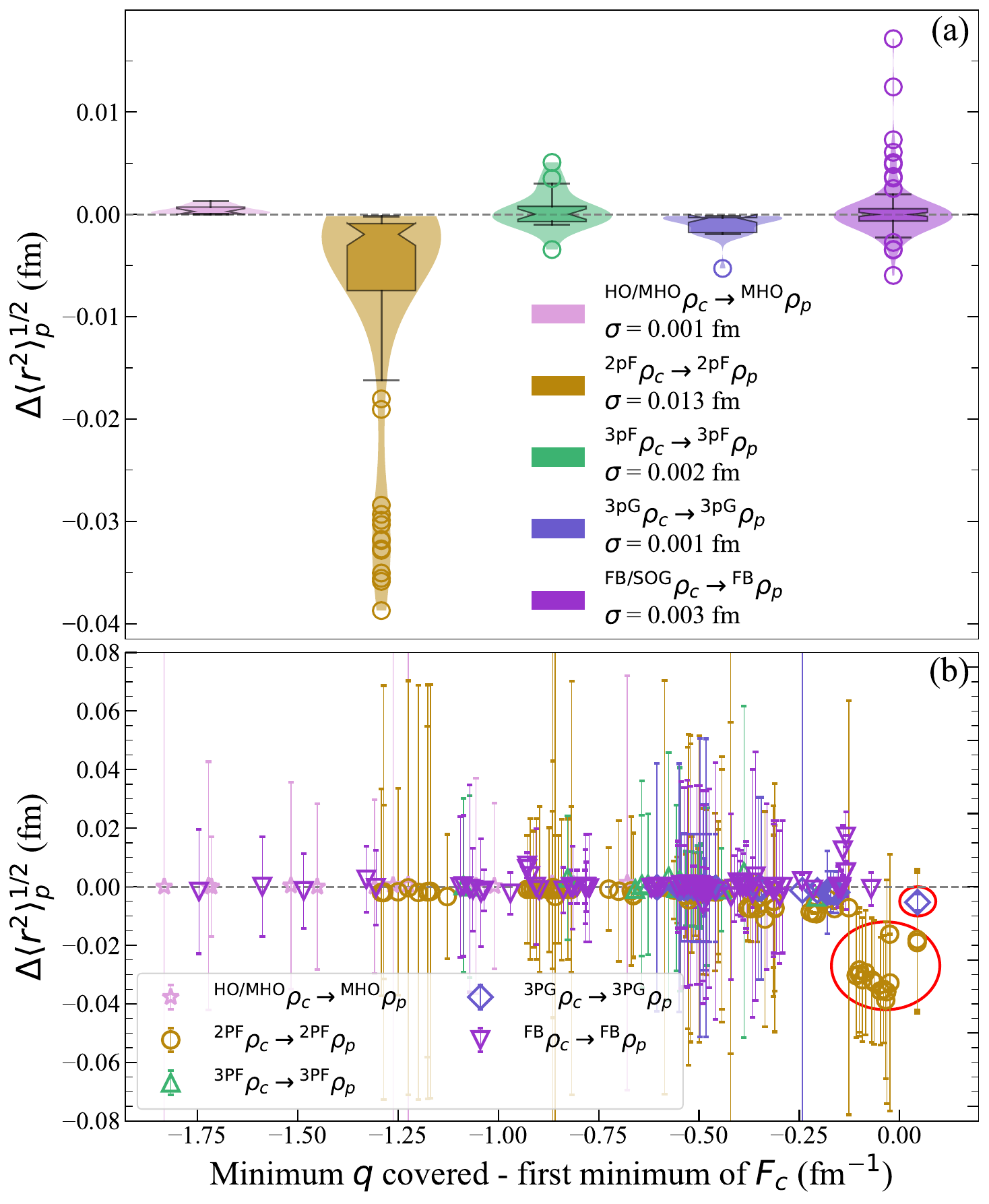}
\caption{(a) Violin and box plots on the radius differences $\Delta \langle r^2 \rangle_{p}^{1/2}$. The different density distributions are distinguished. The root mean square deviations are also given. See the text for more details. (b) The radius differences $\Delta \langle r^2 \rangle_{p}^{1/2}$ as a function of the minimum $q$ covered in the experiments minus the first minimum of $F_c$. The circled data corresponds to the outliers of 2pF and 3pG in Fig.~\ref{fig:plot_box1} (a). Data listed in Tab.~\ref{tab:wrongdata} are excluded.}
\label{fig:plot_box1}
\end{figure}

% \sout{To demonstrate the accuracy of the obtained $\rho_p$ in describing rms radii,} 
The radius differences $\Delta \langle r^2 \rangle_{p}^{1/2}$ 
%$=$$^{2}\langle r^2 \rangle^{1/2}_p$-$^{1}\langle r^2 \rangle^{1/2}_p$, 
between the rms point-proton radii from the derived $\rho_p$ (hereafter as $^{2}\langle r^2 \rangle_p^{1/2}$) and that from Eq.~\eqref{rch} (hereafter as $^{1}\langle r^2 \rangle_p^{1/2}$), are presented as violin and box plots in Fig.~\ref{fig:plot_box1} (a). 
% \sout{Here $\delta\langle r^2 \rangle_{p}^{1/2}$=$(^{from \rho_p}\langle r^2 \rangle_{p}^{1/2} - ^{from \langle r^2 \rangle_{c}}\langle r^2 \rangle_{p}^{1/2})/^{from \langle r^2 \rangle_{c}}\langle r^2 \rangle_{p}^{1/2}$.}
The results are categorized based on different types of distribution functions. The box represents the range in which the middle 50\% of the data lies. The lower and upper edge of the box corresponds to the first quartile (Q1) and the third quartile (Q3), respectively. The area between Q1 and Q3 is the interquartile range, IQR $=\mathrm{Q}3-\mathrm{Q}1$. The groove inside the box indicates the median value. The T-shaped whiskers extend to the furthest data points within 1.5 times the interquartile range from Q1 to Q3. Data outside this range are considered extreme values (outliers) and plotted as empty circles. The violin plots display the data density at different values, reflecting the overall distribution shape and making it easier to visualize the concentration and distribution characteristics of the data. Taking 2pF data (the second box in Fig.~\ref{fig:plot_box1} (a)) as an example, the median value of the data set is approximately 0.002 fm. 
Two whiskers encompass data agrees within 0.018 fm. Outliers outside the lower whisker are identified by hollow circles with 0.018 to 0.04 fm deviations. 

It is observed that MHO achieves an accuracy of 0.002 fm, demonstrating its reliability and applicability in describing the density distribution of nuclei with $Z$ $\leq$ 8. For heavier nuclei, the most commonly applied 2pF function achieves radius accuracy within 0.002 fm except for the marked outliers. With an additional inner depth parameter in Eqs.~\eqref{3PF} and~\eqref{3PG}, 3pF and 3pG data exhibit significantly narrower and more compact deviation distributions, reaching the level of 0.004 fm. They also have fewer outliers recognized than the 2pF case.
% All the results deviated from the typical experimental errors, except for the outliers.

The first dip of the form factor needs to be covered in experiments to determine the radius information~\cite{Suda:2023msg}. As depicted in Fig.~\ref{fig:plot_range} (a) and (d), the relatively low $q$ region, sensitive to the surface area, is crucial for determining rms radii. The deviations of rms radii are illustrated as a function of the minimum $q$ ($q_{\mathrm{min}}$) reached in experiments, minus the first minimum of $F_{c}$ in Fig.~\ref{fig:plot_box1} (b). The radius deviations $\Delta \langle r^2 \rangle_{p}^{1/2} $ increases considerably when $q_{\mathrm{min}}$ is close to the first minimum of $F_{c}$. For 2pF cases, the deviations reach 0.04 fm, while around 0.005 fm for 3pG. These outliers in the box plots in Fig.~\ref{fig:plot_box1} (a) are also marked with red circles in Fig.~\ref{fig:plot_box1} (b) and enumerated in Tab.~\ref{tab: outliers}. Most of data predominantly come from the same study~\cite{sh78a} with relatively higher $q_{\mathrm{min}}$ (close to 1 fm$^{-1}$). Most of these data have deviations outside the experimental uncertainties.
% indicating that the precision of the $\langle r^2 \rangle_{p}^{1/2}$ obtained within the current $q_\mathrm{range}$ is insufficient to match the accuracy of the experimental measurements.
These data also deviate from others and should be used with caution. Taking $^{54}\mathrm{Cr}$ as an example, it gives $\langle r^2 \rangle_{c}^{1/2}$ 3.776(15) fm, while the other two results give 3.673(14) and 3.689(4) fm for which both $q_{\mathrm{min}}$ values 0.15 fm$^{-1}$.
% \textcolor{red}{Taking $^{54}\mathrm{Fe}$ as an example, the outlier data gives $\langle r^2 \rangle_{c}^{1/2}$ 3.732(15) fm, while the other two existing results give 3.680(13) and 3.675(17) fm for which $q_{\mathrm{min}}$ values 0.51 and 0.15 fm$^{-1}$.} 
As for the three outliers marked in 3pF cases, no pronounced correlation is observed. These anomalies could be due to experimental errors or inherent biases since they are measured with a suitable $q_\mathrm{range}$. 

% The derived $\rho_{p}$ can also estimate the uncertainty induced by model-dependent density distributions.
In this work, $F_c$ obtained from $\rho_{c}$ is considered as experimental data to deduce $\rho_{p}$, which has the same $q_\mathrm{range}$ as the experiment covered. The point-proton density $\rho_{p}$ with predefined distribution functions cannot fully capture $F_c$ shape within experiment $q_\mathrm{range}$.
% $\rho_{p}$ with predefined distribution functions cannot fully capture $F_c$ shape within experiment $q_\mathrm{range}$. 
At the same time, $F_c$ information beyond $q_\mathrm{range}$ is lacking. The two reasons above lead to the radii discrepancies between $^{1} \langle r^2 \rangle_{p}^{1/2}$ and $^{2} \langle r^2 \rangle_{p}^{1/2}$. We conclude that if $q_{\mathrm{min}}$ is far below the first dip of $F_{c}$, the accuracy is overall 0.01 fm for 2pF, 0.002 fm for MHO (for $Z \leq 8$ nuclei) and 0.004 fm for 3pF, and 3pG. If one aims for a radius precision of 0.01 fm in the experiment, 2pF with an adequate $q_\mathrm{range}$ is sufficient. 

\begin{table}[H]
\caption{\label{tab: outliers}
Outliers of 2pF, 3pG and FB marked in Fig.~\ref{fig:plot_box1} (a). The model of $\rho_{c}$, minimum $q$ reached in experiments, and the radius difference $\Delta \langle r^2 \rangle_{p}^{1/2}$ are listed in the second, third, and fourth columns, respectively. 
% The ratio of the experimental $\langle r^2 \rangle_{c}^{1/2}$ error \textcolor{red}{????did u use the sqrt?} to $\lvert \delta \langle r^2 \rangle_{p}^{1/2} \rvert$ \textcolor{red}{?????is this right?} $\delta_{\mathrm{exp}}$, is given in the last column.
}
% \textcolor{blue}{The sources of these data are listed in the last column}.}
\renewcommand{\arraystretch}{0.8}
\begin{ruledtabular}
\begin{tabular}{cccc}
\textrm{Nucleus}&
\textrm{Model}&
\parbox{2.3cm}{\textrm{Minimum $q$} \\ \textrm{(fm$^{-1}$)}}&
\parbox{1.5cm}{$\Delta \langle r^2 \rangle_{p}^{1/2}$ \\(fm)}\\
% \parbox{1.5cm}{$\delta_{\mathrm{exp}}$}\\
% \textrm{Reference}\\  
\colrule
$^{50}\mathrm{Cr}$&2pF&0.97&-0.030(17)\\%&\cite{sh78a}\\
$^{52}\mathrm{Cr}$&2pF&0.97&-0.028(17)\\%&\cite{sh78a}\\
$^{53}\mathrm{Cr}$&2pF&0.97&-0.030(17)\\%&\cite{sh78a} \\
$^{54}\mathrm{Cr}$&2pF&0.97&-0.032(17)\\%&\cite{sh78a} \\
$^{54}\mathrm{Fe}$&2pF&0.97&-0.029(18)\\%&\cite{sh78a}\\
$^{56}\mathrm{Fe}$&2pF&0.97&-0.032(17)\\%&\cite{sh78a}\\
$^{58}\mathrm{Fe}$&2pF&1.02&-0.039(22)\\%&Not found\\
$^{59}\mathrm{Co}$&2pF&0.96&-0.033(21)\\%&\cite{sh78a} \\
$^{63}\mathrm{Cu}$&2pF&0.96&-0.036(17)\\%&\cite{sh78a}\\
$^{65}\mathrm{Cu}$&2pF&0.96&-0.035(17)\\%&\cite{sh78a}\\
$^{66}\mathrm{Zn}$&2pF&0.96&-0.036(35)\\%&\cite{li73}\\
$^{68}\mathrm{Zn}$&2pF&0.96&-0.033(41)\\%&\cite{li73}\\
$^{118}\mathrm{Sn}$&2pF&0.84&-0.019(24)\\%&\cite{li73}\\
$^{124}\mathrm{Sn}$&2pF&0.84&-0.018(24)\\%&\cite{li73}\\
$^{209}\mathrm{Bi}$&3pG&0.70&-0.005(31)\\%&\cite{si73b}\\
$^{16}\mathrm{O}$&FB&0.29&0.003(11)\\%&unpublished \\
$^{50}\mathrm{Cr}$&FB&0.15&0.006(5)\\%& \cite{FB505254Cr}\\
$^{52}\mathrm{Cr}$&FB&0.15&0.007(4)\\%&\cite{FB505254Cr} \\
$^{54}\mathrm{Cr}$&FB&0.15&0.004(5)\\%&\cite{FB505254Cr}\\
$^{64}\mathrm{Ni}$&FB&0.51&-0.006(34)\\%
$^{66}\mathrm{Zn}$&FB&0.51&-0.003(36)\\%& unpublished \\
$^{98}\mathrm{Mo}$&FB&0.56&-0.003(22)\\%& \cite{FB98Mo}\\
$^{144}\mathrm{Sm}$&FB&0.60&0.005(12)\\% &\cite{FB144148150152Sm}\\
$^{148}\mathrm{Sm}$&FB&0.60&0.005(8)\\%&\cite{FB144148150152Sm} \\
$^{150}\mathrm{Sm}$&FB&0.60&0.012(8)\\%&\cite{FB144148150152Sm}\\
$^{152}\mathrm{Sm}$&FB&0.60&0.017(8)\\%&\cite{FB144148150152Sm}\\
$^{166}\mathrm{Er}$&FB&0.29&-0.003(28)\\%&\cite{FB166Er}\\
$^{174}\mathrm{Yb}$&FB&0.32&0.004(42)\\%&unpublished \\
\end{tabular}
\end{ruledtabular}
\end{table}

\begin{figure}[htpb]
\centering
\includegraphics[width=0.48\textwidth]{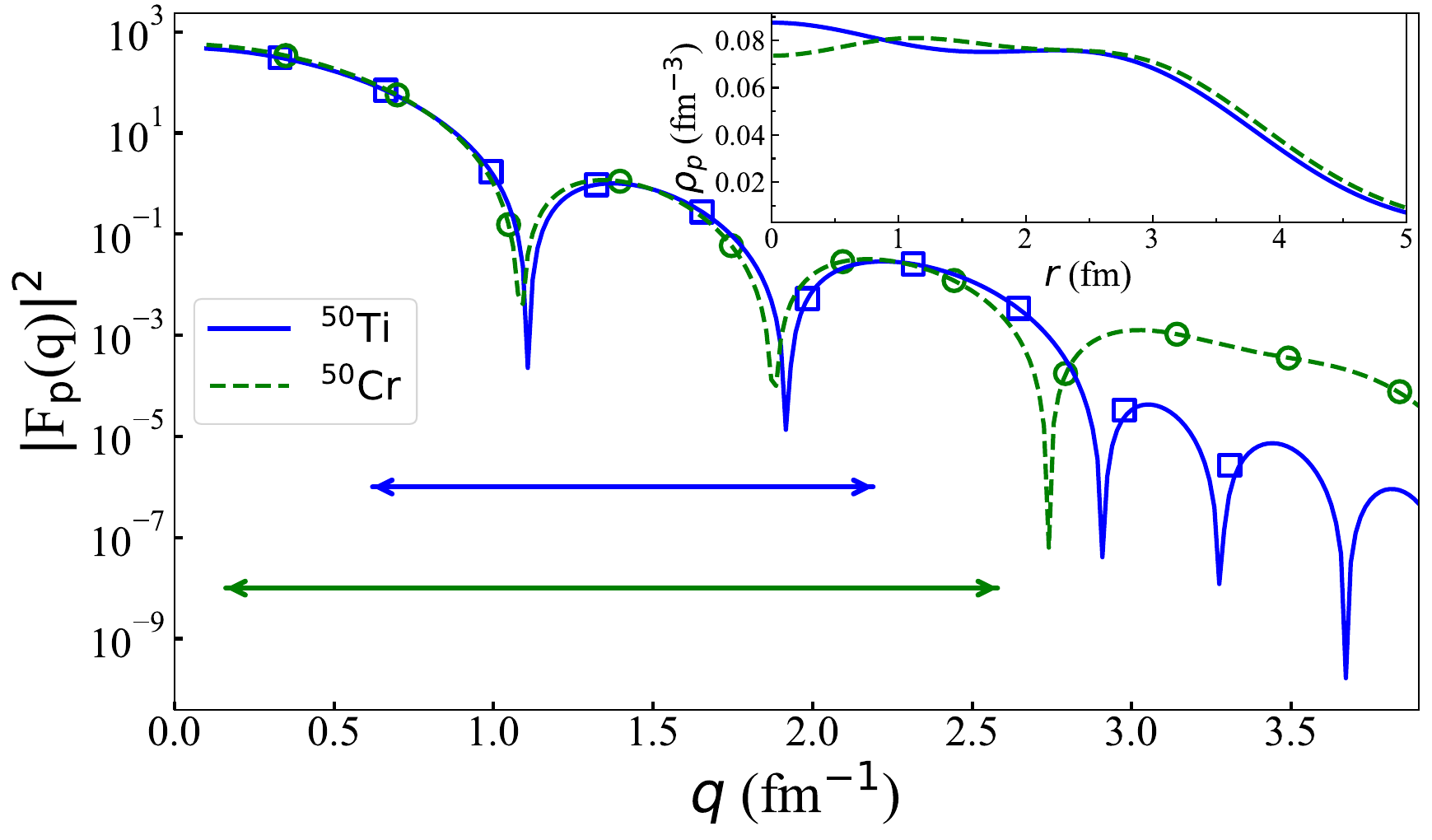}
\caption{The deduced $F_p$ of $^{50}\mathrm{Ti}$ and $^{50}\mathrm{Cr}$ in the FB form with blue solid and green dashed lines
% correspond to the cases with $\Delta \langle r^2 \rangle_{p}^{1/2}$ of 0.017(8) and 0.002(11) fm, 
respectively. The points are FB coefficients. Arrows represent $q_{\mathrm{range}}$ covered in experiments. The inset is the corresponding $\rho_p$.
}
\label{fig:plot_FB_formfactor}
\end{figure}

The outliers in FB cases in Fig.~\ref{fig:plot_box1}(a) are listed in Tab.~\ref{tab: outliers}. Overall, most of the outliers fall within the error bars, and they exhibit no obvious correlation with $q_{\mathrm{min}}$, as demonstrated in Fig.~\ref{fig:plot_box1} (b). % but probably due to the FB uncertainties.
We take $^{50}\mathrm{Ti}$ and $^{50}\mathrm{Cr}$ as examples. Their radius differences are -0.002(3) and 0.006(6) fm (see Tab~\ref{tab:grossdatafb} in Appendix for details). Their deduced $F_p$, and the relevant densities $\rho_p$ are compared in Fig.~\ref{fig:plot_FB_formfactor}.
% Their experimental $F_c$, deduced $F_p$, and the relevant densities $\rho_c$ and $\rho_p$ are shown.
Two $F_p$ start to deviate considerably from each other when $q$ exceeds 3 fm$^{-1}$, reflecting the uncertainties caused by extrapolation beyond experimental $q_{range}$. The differences at high $q$ are accordingly reflected in the central density distributions. As shown by the inset, a significant difference between the two density distributions in the center region is observed.
% \iffalse 
% The blue one keeps the same peak shape up to 3 fm$^{-1}$, with narrower peaks observed between 3 and 3.5 fm$^{-1}$, outside the data points. Correspondingly, the $F_p$ shows only a higher amplitude, without any noticeable changes in the dips. The radius deviation is only 0.03\%. In the inset, $\rho_p$ presents a minimal difference from $\rho_c$ except for a slight difference at the shoulder range. However, the green one exhibits narrower peaks after 2 fm$^{-1}$, with the most noticeable change occurring between 3 and 3.5 fm$^{-1}$. The deduced $F_p$ does not closely follow the behavior of the $F_c$. It fails to show the dip at around 3.2 fm$^{-1}$, leading to a larger radius deviation, 0.34\%. Accordingly, $\rho_p$ shows more evident oscillations than $\rho_c$, completely different from the other one. 
% \fi
Therefore, caution should be taken when utilizing the central density pattern and form factor at high $q$ for the few data with large $\Delta \langle r^2 \rangle_{p}^{1/2}$ values.

\subsection{Two-parameter Fermi point-proton densities}
\begin{figure}[!htpb]
\centering
\includegraphics[width=0.35\textwidth]{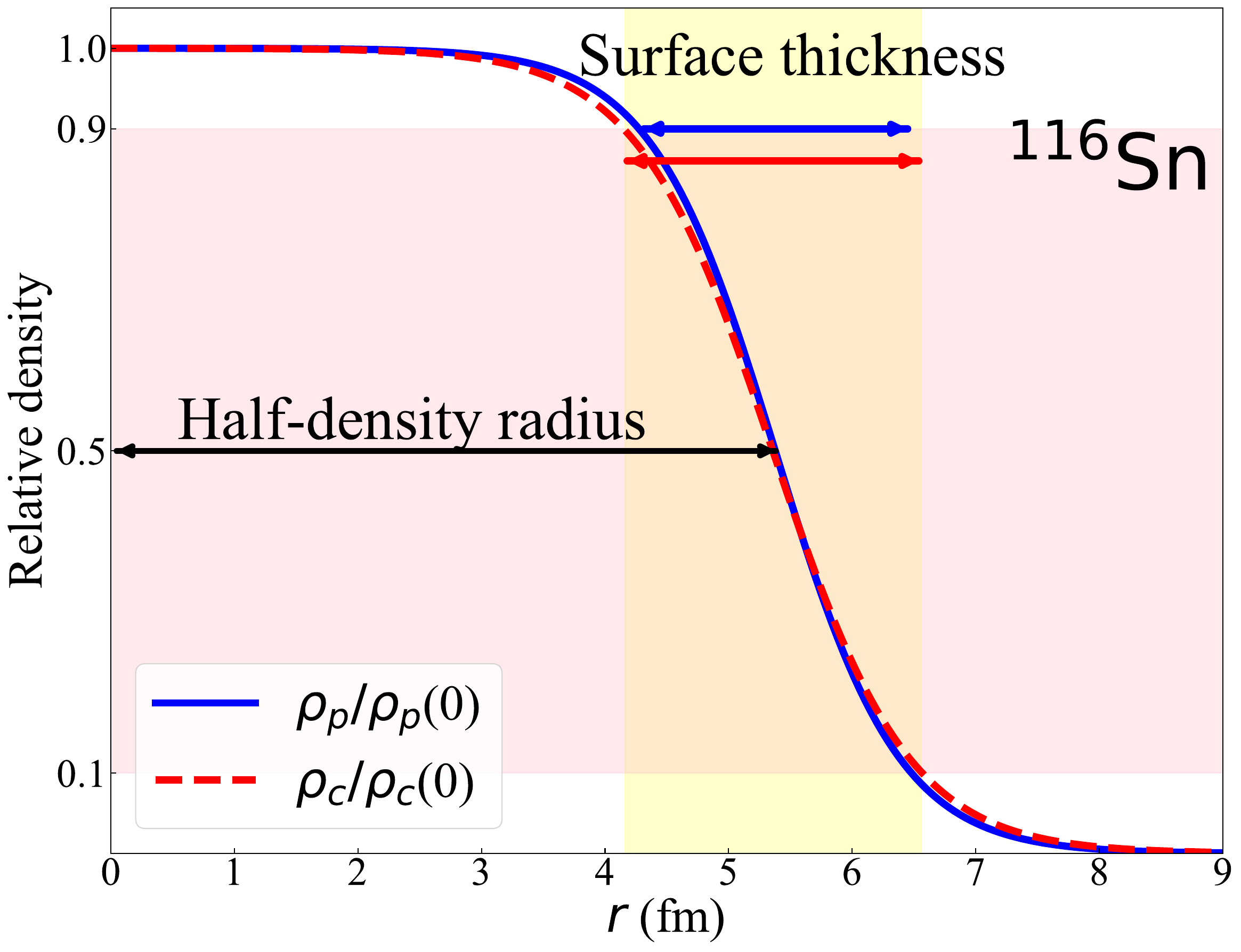}
\caption{\label{fig:charge_and_proton_density}
  Charge density $\rho_c$ (red dashed line) and point-proton density $\rho_p$ (blue solid line) of $^{116}\mathrm{Sn}$ in the 2pF form. The density is scaled by dividing the central density $\rho$(0). The half-density radius and surface thickness are also indicated.}
\end{figure}

\iffalse
Fig.~\ref{fig:form_factor_and_density} (a) describes the $^{116}\mathrm{Sn}$ 2PF charge form factor, deduced proton form factor $F_{p}$, and the neutron form factor $F_{n}$ multiply Sachs form factor $G_{En}$ and times 1000. 
\fi
For nuclei with $Z > 8$, most of the data have been given in the 2pF distribution~\cite{DEVRIES1987495}:
\begin{equation}\label{2pf2}
\rho_{}(r)=\frac{\rho_0}{1+e^{\frac{r-c}{z}}} \;,
\end{equation}
where $\rho_0$, $c$, and $z$ are the normalization coefficient, half-density radius, and surface diffuseness, respectively. This allows a systematic comparison of the density shapes and parameters between $\rho_c$ and $\rho_p$. 
Taking $^{116}\mathrm{Sn}$ as an example, its $\rho_p$ and 
$\rho_c$ are shown in Fig.~\ref{fig:charge_and_proton_density}.
%shows c and $\rho_p$ of $^{116}\mathrm{Sn}$ in 2PF form. Compared to $\rho_c$, $\rho_p$ exhibits a slightly narrower and steeper surface diffuseness region. 
Although having similar distributions, the surface diffuseness parameter $z$ decreases slightly from 0.550(9) fm in $\rho_c$ to 0.502(5) fm in $\rho_p$, while the radius parameter $c$ remains almost unchanged, merely varying from 5.358(22) to 5.377(21) fm. 
%The decrease in the surface parameter and the nearly constant radius parameter primarily come from the numerically larger shape of $F_p$ compared to that of $F_c$.
%The proton electric form factor $G_{Ep}$ has a monotonically decreasing trend with its maximum at $G_{Ep}(0)=1$~\cite{universe9040182}. Namely, the proton form factor $F_{p}$ yields a higher value than the charge form factor $F_{c}$ according to Eq.~\eqref{fourier}. For the neutron term in Eq.~\eqref{fourier}, a little decline induced is a natural consequence but is hardly discerned due to $G_{En}$ being hundreds of times smaller than the $G_{Ep}$. It is recognized that the radius and the surface diffuseness parameters independently control the dip position and the diffraction maximum height of the form factor~\cite{sudaPPNP}, same as the hadron probe case~\cite{Hatakeyama2018}. 
% \textcolor{brown}{The deduced point-proton density shows a sharper surface thickness marked by the blue arrow compared to the red one. The almost no dip position shift of $F_p$ leads to no significant variation for the radius parameter, as indicated by the black arrow at the half-density position. The main distinction between $\rho_{c}$ and $\rho_p$ arises from the contribution of the intrinsic proton and neutron densities. This results in a 2\% ???? more compact spatial point-proton distribution in coordinate space, while the radius parameters remain nearly identical.}
% \textcolor{red}{discuss a bit on the diffuseness: derecsing., then Fig. c shows the parameter comparisons: what you see, which correlations do you find, its potential application. taking 28Si as an example, how to deduce $\rho_p$ parameters from $\rho_c$. Then, discuss the outliers. }

\begin{figure}[!htpb]
\centering
\includegraphics[width=0.48\textwidth]{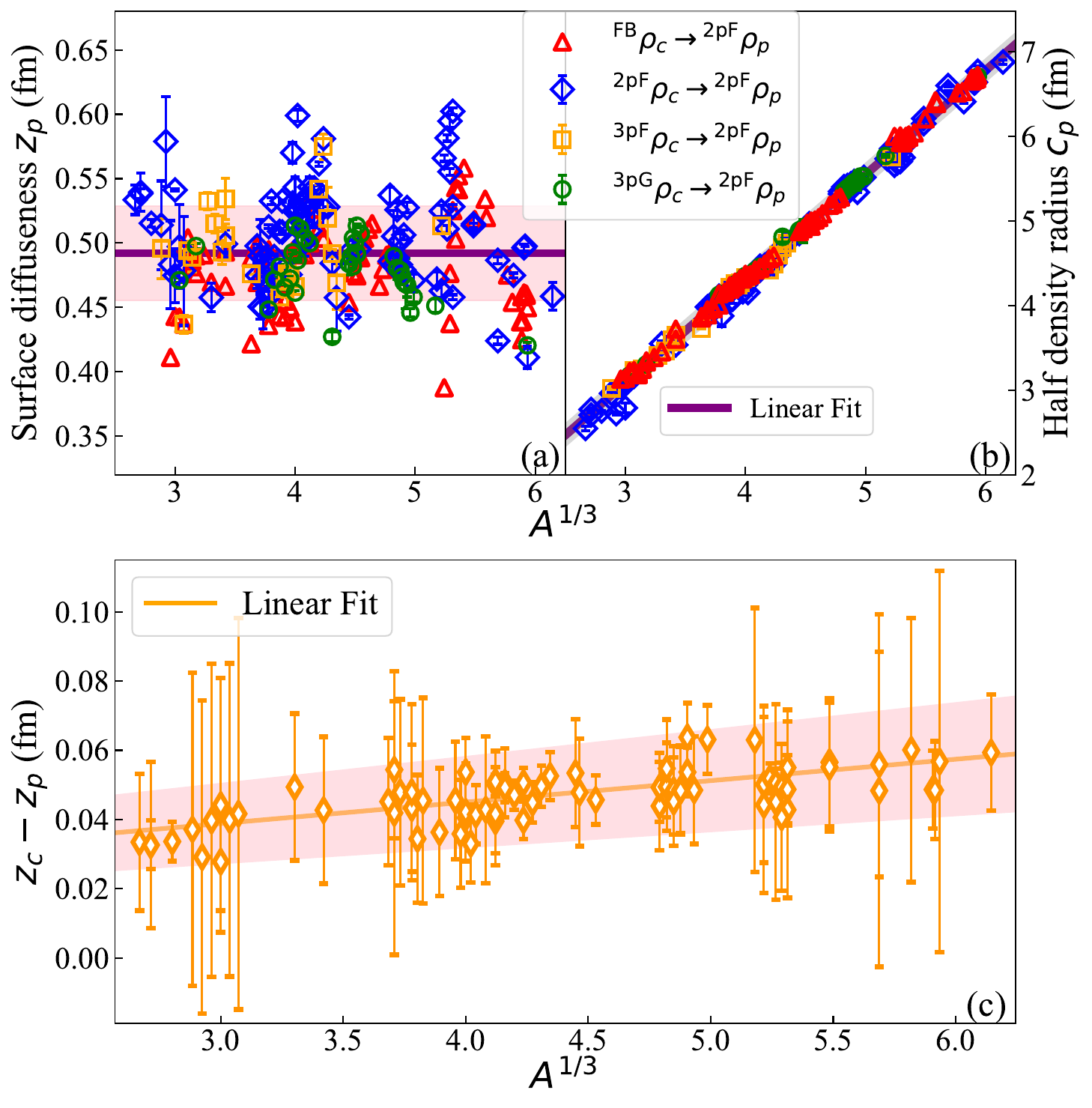}
\caption{\label{fig:plot_rate_plus_emprical} (a) Surface diffuseness $z_p$ and (b) half-density radius $c_p$ of point-proton density $\rho_p$ in 2pF form from four kinds of $\rho_c$ as a function of $A^{1/3}$. The solid line in (a) is the average value and the colored region represents the standard deviation. (c) Difference in surface diffuseness parameters between $\rho_c$ and $\rho_p$, $z_c-z_p$, as a function of $A^{1/3}$. The solid line illustrates the linear fit of $z_c-z_p$. The colored region depicts the loci of the regression's 95\% confidence bands.}
\end{figure} 

The deduced $\rho_p$ parameters in 2pF form from $\rho_c$ in 2pF, 3pF, 3pG and FB forms, are presented in Fig.~\ref{fig:plot_rate_plus_emprical} (a) and (b). Data listed in Tabs.~\ref{tab:wrongdata} and~\ref{tab: outliers} are excluded. The surface parameters ($z_p$) exhibit an average value of 0.49 fm with a standard deviation of 0.037 fm. It is also evident that some 2pF data exhibit significant biases. Even for the same nuclei, surface parameters can significantly differ in different experimental results due to different $q_{\mathrm{range}}$. This is also noticed in muonic atom spectroscopy~\cite{XIE2023PLB}.
For example, two existing measurements for $^{146}\mathrm{Nd}$ mainly differing in maximum $q$ reached, namely, 0.73 and 2.94 $\mathrm{fm}^{-1}$, have the surface parameters of 0.6321(30) fm and 0.556(20) fm, while the radius parameters of 5.6541 fm and 5.867(32) fm. However, they give consistent rms radii, namely, 4.970(5) fm and 4.993(37) fm. 

In Fig.~\ref{fig:plot_rate_plus_emprical} (b), the deduced half-density radius parameters ($c_p$) of 2pF $\rho_p$, derived from 2pF, 3pF, 3pG, and FB $\rho_c$, all exhibit a clear adherence to the $A^{1/3}$ rule, namely,
\begin{equation}\label{cemprical}
    c_p=1.232(6)A^{1/3}-0.612(25) \;.
\end{equation}
The linear coefficient of 0.995 indicates the consistency among different types of $\rho_c$ data.
%. Moreover, the linear trends exhibited by the results from 3PF, 3PG, and FB types of $\rho_c$ are even more pronounced than those from the $\rho_c$ in 2PF form. The radius parameters of the $\rho_p$ in 3PF form are likewise examined, but they did not show a strong linear relationship as those of the 2PF.

% \textcolor{brown}{\sout{Gloablly, the radius parameters show less than 1\% difference for all the nuclei except for a bit of fluctuation for light nuclei with the mass number around 20. The variations are comparable to the relevant experimental uncertainty. %The changes are negligible compared to the value of the radius parameter itself.
% }}

Figure~\ref{fig:plot_rate_plus_emprical} (c) presents the difference between $\rho_c$ and $\rho_p$ surface diffuseness parameters as a function of $A^{1/3}$.
%The main change from $\rho_c$ to $\rho_p$ lies in reducing the surface diffuseness region. As seen in Fig.~\ref{fig:plot_rate_plus_emprical} (a), the relative variation of the surface diffuseness parameter $\delta z$, $(z^p-z^c)/z^c$, 
The difference is much larger than the experiment uncertainties (typically 0.019 fm). 
% \textcolor{red}{As indicated in Eq.~\eqref{fourier}, the difference is the summing effect of both single-proton and single-neutron charge density distributions, which have opposite signs. Overall, the dominant contribution is single-proton charge density.} 
The difference increases smoothly with $c_c$, indicating the contribution of single-proton charge density is more significant as nuclei become heavier.
%To further delineate the relative difference of surface parameters, results neglecting neutron term in Eq.~\eqref{fourier} are also deduced, in which the variation of surface parameters entirely comes from single-proton charge density. As observed in more diffuse surface regions in $\rho_c$, single-proton charge density reduces the surface parameter around 7\% to 15\% in deducing $\rho_p$ as the mass number increases from 20 to over 200. The slightly negatively charged $\rho_n$ results in an expansion of the $\rho_p$ surface diffuseness region. This effect only contributes 1\% to 2\% in light nuclei ($A$ around 20), but up to 5\% for heavy nuclei ($A$ over 200). 
%and Their effects on the surface diffuseness parameter are opposite, resulting in the total relative difference in surface diffuseness about 5\% to 10\% with increasing mass number, which is significantly larger than the uncertainties of $z^c$, of typically of 2.5\%. 
The following empirical formula can describe the systemic difference in surface parameters:
\begin{equation}\label{zemprical}
    z_p=z_c-(0.0062(8)A^{1/3}+0.020(4)) \;.
\end{equation}
% \textcolor{red}{this formula is practically useless since one has to know $C_c$ first. Combining Eq.8 and 9 may give a relationship as a function of $A^{1/3}$!}
% Differences in surface parameters between $\rho_p$ and $\rho_{c}$ show a slight linear increase with the increase of $A^{1/3}$ and can be parametrized by a least square fit: $z^{c}=z^{p}+0.0059(7)A^{1/3}+0.020(3)$. Approximately a 0.05 fm difference for all the nuclei (more appliable for nuclei with proton number larger than 30). This is equivalent to a 10\% relative reduction (for an average value of 0.49 fm shown in Fig.~\ref{fig:plot_R_para} (a)). 
 % \sout{The hollow data points in Fig.~\ref{fig:form_factor_and_density} (b) with significant systematic deviations are due to inaccuracies caused by insufficient experiment $q$ measurements. These will be discussed in the next section.}
Although the linear coefficient of 0.5 indicates a moderate correlation, this relationship still captures the general variation trend of the diffuseness parameters. A moderate correlation may arise from inconsistencies in surface parameters across the results of different experiments. Eqs.~\eqref{zemprical} and~\eqref{cemprical} offer an empirical way to determine $\rho_p$ from $\rho_{c}$ in 2pF case, and vice versa. The formulas will be useful for a quick estimation of $\rho_p$.
%For example, these two equations yield radius and surface parameters 5.40(4) and 0.50(1) $fm$ for $^{116}\mathrm{Sn}$, consistent with 5.38(2) and 0.502(5) $fm$ from deduced $\rho_p$. Corresponding $rms$ radii is 4.58(4) and 4.57(2) $fm$. 
\iffalse
\textcolor{brown}{Due to the simplistic consideration of $\rho_n$, examining the parameters' sensitivity to $\rho_n$ in Eq.~\eqref{fourier} is also conducted. Fig.~\ref{fig:plot_R_para}(c) illustrates the relative deviation of the parameters variation with and without the inclusion of $\rho_n$ in Eq.~\eqref{fourier}. It is observed that for all the nuclei, the variation is quite minimal for the radius parameter, approximately within the range of 0.1\%-0.2\%. For the surface diffuseness parameter, the contribution of $\rho_n$ leads to a variation of 1.5\% to 4.5\% with increasing mass number.}
\fi

\subsection{Point-proton densities for $^{48}\mathrm{Ca}$ and $^{208}\mathrm{Pb}$}

Neutron skin thickness of $^{48}\mathrm{Ca}$ and $^{208}\mathrm{Pb}$ serve as ideal targets for investigating the equation of state of nuclear matter. Their relevant point-proton density distributions are fundamentally important as inputs for the analysis of neutron density distribution and neutron-skin thickness.  

\begin{figure}[!htpb]
\centering
\includegraphics[width=0.5\textwidth]{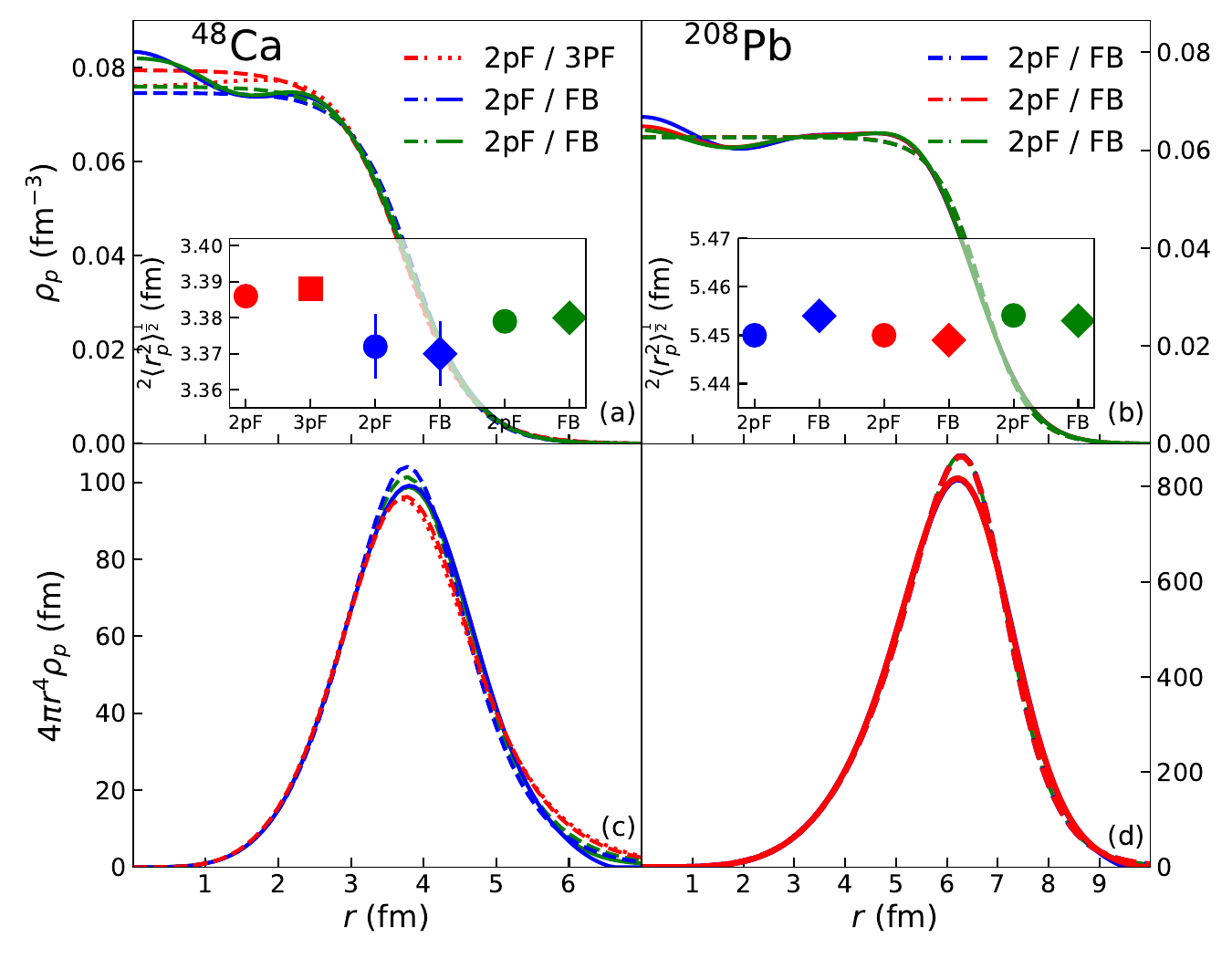}
\caption{\label{fig:plot_48Ca208Pb} The deduced $\rho_p$ of $^{48}\mathrm{Ca}$ (a) and $^{208}\mathrm{Pb}$ (b) and their relevant radial integrated (c) and (d) in the FB form (solid line), the 2pF form (dashed line), and the 3pF form (dotted line). The inset is the corresponding $^{2}\langle r^2 \rangle_{p}^{1/2}$ with 2pF, 3pF, and FB in circle, square, and diamond, respectively.}
\end{figure} 

There are three experimental $\rho_c$ results for $^{48}\mathrm{Ca}$, which are presented in the 3pF, FB, and SOG forms, respectively. Their corresponding $^{1}\langle r^2 \rangle_{p}^{1/2}$ are 3.388, 3.370(9) and 3.379 fm. $\rho_p$ in two different density forms can be deduced for each $\rho_c$ form. For example, for $\rho_c$ in the FB form, we have given $\rho_p$ in 2PF and FB forms.  
For $^{208}\mathrm{Pb}$, there are three experimental $\rho_c$ results, two of which are in the FB forms and the other in the SOG form. They give $^{1}\langle r^2 \rangle_{p}^{1/2}$ of 5.454(2), 5.450(1) and 5.454(2) fm, respectively. In total, six different $\rho_p$ and $^{2}\langle r^2 \rangle_{p}^{1/2}$ are presented for both $^{48}\mathrm{Ca}$ and $^{208}\mathrm{Pb}$. 
The details can be found in Tables.~\ref{tab:grossdata} and~\ref{tab:grossdatafb}.  

In Fig.~\ref{fig:plot_48Ca208Pb} (a) and (b), we present the deduced $\rho_p$ in various density forms for $^{48}\mathrm{Ca}$ and $^{208}\mathrm{Pb}$. 
 Similar to the case in Fig.~\ref{fig:plot_range}, the 2pF, 3pF and FB cases exhibit similar behavior at the tail, while more details are provided in FB form at the central range than a completely flat distribution given by 2pF. The 2pF, 3pF, and FB cases provide almost identical $^{2}\langle r^2 \rangle_{p}^{1/2}$, indicating the consistency of different density distributions in describing radii. The radial integrand functions are presented in Fig.~\ref{fig:plot_48Ca208Pb} (c) and (d). The interior oscillations in the volume integral are negligible, while the most significant differences occur at the maximum value of the radial integral. In all, the 2pF and 3pF $\rho_p$ provide sufficient precision for the radius, but FB offers more detailed density. 

\section{summary}\label{sec4}
In this paper, we deduce the point-proton density distributions of 130 stable nuclei from $^{7}\mathrm{Li}$ to $^{232}\mathrm{Th}$ by unfolding $\rho_c$ measured in elastic electron scattering. The $\langle r^2 \rangle_{p}^{1/2}$ values calculated from the derived $\rho_p$ agree in 0.2\% with the one directly obtained from $\langle r^2 \rangle_{c}^{1/2}$ in model-dependent forms. This is smaller than the experimental radius accuracy. We have identified cases exhibiting large $\langle r^2 \rangle_{p}^{1/2}$ deviations due to the limited $q$ coverage in experiments or inconsistent scattering experiments. Caution is advised when using these data. In the 2pF cases, the surface diffuseness parameter of $\rho_p$ shows a reduction of about 10\% compared to that of $\rho_c$. The transition from $\rho_c$ to $\rho_p$ for the surface parameter follows a linear trend with $A^{1/3}$, whereas the half-density radius parameter remains almost unchanged and follows well the $A^{1/3}$ rule. Finally, we show the $\rho_p$ for $^{48}\mathrm{Ca}$ and $^{208}\mathrm{Pb}$, the two important nuclei for studying nuclear equation of state.
% The commonly used model-independent FB distribution does not always possess the most accurate $rms$ radius information. The inaccurate data is filtered out through the box plot, they present abnormal $F_c$ shapes at the high $q$ range, even under the experiment measurement range. Beware of the central area density distribution behavior provided by FB. Except for a few data, most FB data provides an accuracy of 0.1\% or even higher.
% the choice of $R_{cut}$ and the number of constants significantly impact radius accuracy. An excessively small $R_{cut}$ may fail to adequately describe the tail density. In contrast, an overly large $R_{cut}$ necessitates more constants to cover the high momentum transfer region, which can lead to the need for additional experimental points. Therefore, $R_{cut}$ should be chosen appropriately, and extrapolation of the shape factor from experimental measurements is necessary to describe radius information accurately.
The derived point-proton densities can be used as inputs for nuclear reaction models and assessments of nuclear structure models. Moreover, they can be benchmarks when extracting radii in charge-changing cross-section measurements of unstable nuclei and a basis for analyzing experimental data to extract neutron skin.

\begin{acknowledgments}
Discussions with S. Terashima and J. Li are gratefully acknowledged. 
This work was partly supported by the National Natural Science Foundation of China (Nos. 12325506, 11961141004, 11922501, 12475119) and the 111 Center (Grant No. B20065). 
\end{acknowledgments}

%\bibliography{Reference}

\newpage
\section*{Model-dependent and model-independent densities distribution}\label{app1}

The nuclear charge density distributions are usually written in parameterized functions, including HO, MHO, 2pF, 3pF, 3pG, SOG, and FB.  

The model-dependent Harmonic-oscillator function is written as~\cite{DEVRIES1987495}:
\begin{equation}
\rho_{}(r)=\rho_0(1+\alpha(\frac{r}{a})^2)e^{-(\frac{r}{a})^2},
\label{ho}
\end{equation}
where $\alpha=\frac{\alpha_0a_0^2}{a^2+\frac{3}{2}\alpha_0(a^2-a_0^2)}$, $a_0^2=\frac{A(a^2-a_p^2)}{A-1}$, $\alpha_0=\frac{Z-2}{3}$, $a_p^2=\frac{2}{3} r^2_{p}$. The modified harmonic-oscillator model has the same expression as HO but with $\alpha$ as an additional free parameter. 
% {for each case, you can briefly comment on its application. e.g. HO/MHO is widely used for its simplicity for $Z \le 8$. }

The model-dependent two-parameter Fermi function is written as:
\begin{equation}\label{2pf}
\rho_{}(r)=\rho_0\frac{1}{1+e^{\frac{r-c}{z}}},
\end{equation}
where $c$ and $z$ correspond to the half-density and the surface diffuseness parameter, respectively.

The model-dependent three-parameter Fermi function is written as:
\begin{equation}\label{3PF}
\rho_{}(r)=\rho_0\frac{1+w\frac{r^2}{c^2}}{1+e^{\frac{r-c}{z}}},
\end{equation}
where $w$ corresponds to the inner depth parameter.

The model-dependent three-parameter Gaussian function is described as:
\begin{equation}\label{3PG}
\rho_{}(r)=\rho_0\frac{1+w\frac{r^2}{c^2}}{1+e^{\frac{r^2-c^2}{z^2}}}.
\end{equation}

The model-independent Fourier-Bessel series~\cite{Dreher}:
\begin{equation}\label{fb}
\rho_{}(r)=\left\{
\begin{aligned}
    &  \sum_{\nu}^{N_{\text{FB}}}a_{\nu}j_0(\frac{\nu\pi r}{R_{\mathrm{cut}}}), & & \text{for $r \leqslant R_{\mathrm{cut}}$}, \\
    & 0 & & \text{for $r > R_{\mathrm{cut}}$},
\end{aligned}
\right.
\end{equation}
where $j_0(\frac{\nu\pi r}{R_{\mathrm{cut}}})=\frac{\mathrm{sin}(\nu\pi r/R_{\mathrm{cut}})}{\nu\pi r/R_{\mathrm{cut}}}$ denotes the Bessel function of order zero, and $R_{\mathrm{cut}}$ is the cutoff radius of the charge distribution, truncating the expansion after $\nu$ terms and assuming the charge is zero for $r\geqslant R_{\mathrm{cut}}$.
% with the normalization
% \begin{equation}\label{fbnormal}
% 4\pi \int \rho_c(r)r^2dr=4\pi \sum_{\nu=1}^{N_{\text{FB}}}\frac{(-1)^{\nu+1}a_{\nu}R_{cut}^3}{(\nu\pi)^2}=Z
% \end{equation}
The coefficients $a_\nu$ can be determined by the $F_c$,
\begin{equation}\label{fbcoefficients}
a_\nu=\frac{q_\nu^2}{2\pi R_\mathrm{cut}}F_c(q_\nu)~~\text{with}~~q_\nu=\frac{\nu\pi}{R_\mathrm{cut}}.
\end{equation}
The considered number of FB coefficients is related to the maximum value of the momentum transfer $q_{\text{max}}$:
\begin{equation}\label{Nfb}
N_{\text{FB}} =\frac{R_{\mathrm{cut}}q_{\text{max}}}{\pi}.
\end{equation}

The model-independent Sum-of-Gaussians function:
\begin{equation}\label{sog}
\rho_{}(r)=\frac{Z}{2\pi^{\frac{3}{2}}\gamma^3}\sum_{i=1}^{12}\frac{Q_i}{1+\frac{2R_i^2}{\gamma^2}}[e^{-(\frac{r-R_i}{\gamma})^2}+e^{-(\frac{r+R_i}{\gamma})^2}],
\end{equation}
where $R_i$ and $Q_i$ are the position and amplitude of the Gaussians. The value of $Q_i$ indicates the fraction of charge contained in the $i^{\mathrm{th}}$ Gaussian, normalized such that $\sum_{i}Q_i$ = 1, and $\gamma$ is the rms radius of the Gaussians.

\begin{widetext}
\newpage
\section*{Explanation of Tables}
\textbf{Table III. Point-proton density distribution parameters\\}
\begin{tabular}{@{}p{0.0in}p{1.5in}p{4.5in}@{}}
& $\rho_{c}$ & Density distribution function for $\rho_{c}$ \\
& $\langle r^2 \rangle^{1/2}_c$ & rms charge radius  \\
& $^{1}\langle r^2 \rangle^{1/2}_p$ & rms point-proton radius calculated from $\langle r^2 \rangle^{1/2}_c$ \\
& $\rho_{p}$ & Density distribution function adopted for the deduced $\rho_{p}$\\
& $^{2}\langle r^2 \rangle^{1/2}_p$ & rms point-proton radius calculated from the deduced $\rho_{p}$\\
& $c$ or $a$ & The values are given for the parameter $c$ if 2PF, 3PF, or 3PG model has been used and for the parameter $a$ if the MHO model has been used\\
& $z$ or $\alpha$ & The values are given for the parameter $z$ if 2PF, 3PF, or 3PG model has been used and for the parameter $\alpha$ if the MHO model has been used\\
& $w$ & The parameter $w$ of the 3PF and 3PG models\\
& $\Delta \langle r^2 \rangle_{p}^{1/2}$ &  $^{2}\langle r^2 \rangle^{1/2}_p$-$^{1}\langle r^2 \rangle^{1/2}_p$\\
& HO & Harmonic-oscillator model\\
& MHO & Modified harmonic-oscillator model\\
& 2pF & Two-parameter Fermi model\\
& 3pF & Three-parameter Fermi model\\
& 3pG & Three-parameter Gaussian model\\
& FB & Fourier-Bessel expansion\\
\\
\end{tabular}

\newpage
\renewcommand{\arraystretch}{1.2} 
\begin{longtable}{cccccccccc}
\caption{Point-proton density distribution parameters. Units are fm.} \label{tab:grossdata}\\
\hline
\textrm{Nucleus} &
%\parbox{1.0cm}{\centering Model \\ ($\rho_{c}$)}&
\textrm{$\rho_{c}$} &
\parbox{1.0cm}{\centering $\langle r^2 \rangle^{1/2}_c$ \\ (fm)} & 
\parbox{1.0cm}{\centering $^{1}\langle r^2 \rangle^{1/2}_p$ \\ (fm)} &
%\parbox{1.0cm}{\centering Model \\ ($\rho_{p}$)}& 
\textrm{$\rho_{p}$} &
\parbox{1.0cm}{\centering $^{2}\langle r^2 \rangle^{1/2}_p$ \\ (fm)} &
\parbox{1.0cm}{\centering $c$ or $a$ \\ (fm)}&
\parbox{1.2cm}{\centering $z$ or $\alpha$ \\ (fm)} &
\textrm{$w$} &
\parbox{2cm}{\centering $\Delta \langle r^2 \rangle_{p}^{1/2}$ \\ (fm)} \\
\hline
\endfirsthead
\multicolumn{8}{c}%
{\tablename\ \thetable\ -- \textit{(continued)}} \\
\hline
\textrm{Nucleus} &
%\parbox{1.0cm}{\centering Model \\ ($\rho_{c}$)} &
\textrm{$\rho_{c}$} &
\parbox{1.0cm}{\centering $\langle r^2 \rangle^{1/2}_c$ \\ (fm)} & 
\parbox{1.0cm}{\centering $^{1}\langle r^2 \rangle^{1/2}_p$ \\ (fm)} &
%\textrm{$\langle r^2 \rangle^{1/2}_c$ (fm)} &
%\textrm{$^{1}\langle r^2 \rangle^{1/2}_p$ (fm)} &
\textrm{$\rho_{p}$} &
%\parbox{1.0cm}{\centering Model \\ ($\rho_{p}$)} &
\parbox{1.0cm}{\centering $^{2}\langle r^2 \rangle^{1/2}_p$ \\ (fm)} &
%\textrm{$^{2}\langle r^2 \rangle^{1/2}_p$ (fm)} &
\parbox{1.0cm}{\centering $c$ or $a$ \\ (fm)} &
\parbox{1.0cm}{\centering $z$ or $\alpha$ \\ (fm)} &
\textrm{$w$} &
\parbox{2cm}{\centering $\Delta \langle r^2 \rangle_{p}^{1/2}$ \\ (fm)} \\
\hline
\endhead
\hline
\multicolumn{8}{r}{\textit{(continued)}} \\
\endfoot
\hline
\endlastfoot

$^{7}\mathrm{Li}$&HO&2.394(30)&2.274(30)&MHO&2.274(30)&1.656(16)&0.419(45)&\mbox{}&0.00(4)\\
$^{9}\mathrm{Be}$&HO&2.519(13)&2.403(13)&MHO&2.403(14)&1.676(16)&0.832(45)&\mbox{}&0.00(1)\\
$^{9}\mathrm{Be}$&HO&2.495(100)&2.377(100)&MHO&2.377(101)&1.653(16)&0.874(45)&\mbox{}&0.00(14)\\
$^{10}\mathrm{B}$&HO&2.452(140)&2.327(140)&MHO&2.328(143)&1.586(16)&1.261(45)&\mbox{}&0.001(200)\\
$^{11}\mathrm{B}$&HO&2.419(140)&2.297(140)&MHO&2.297(143)&1.570(16)&1.199(45)&\mbox{}&0.001(200)\\
$^{12}\mathrm{C}$&FB&2.478(15)&2.348(15)&\renewcommand{\arraystretch}{0.8}\begin{tabular}{@{}c@{}}MHO\\FB\end{tabular}&\renewcommand{\arraystretch}{0.8}\begin{tabular}{@{}c@{}}2.354(15)\\2.346(15)\end{tabular}&\renewcommand{\arraystretch}{0.8}\begin{tabular}{@{}c@{}}1.605\\\mbox{} \end{tabular}&\renewcommand{\arraystretch}{0.8}\begin{tabular}{@{}c@{}}1.164\\\mbox{} \end{tabular}&\renewcommand{\arraystretch}{0.8}\begin{tabular}{@{}c@{}}\mbox{}\\\end{tabular}&\renewcommand{\arraystretch}{0.8}\begin{tabular}{@{}c@{}}0.007(21)\\-0.002(21)\end{tabular}\\
$^{12}\mathrm{C}$&FB&2.464(12)&2.339(12)&\renewcommand{\arraystretch}{0.8}\begin{tabular}{@{}c@{}}MHO\\FB\end{tabular}&\renewcommand{\arraystretch}{0.8}\begin{tabular}{@{}c@{}}2.339(12)\\2.339(12)\end{tabular}&\renewcommand{\arraystretch}{0.8}\begin{tabular}{@{}c@{}}1.594\\\mbox{} \end{tabular}&\renewcommand{\arraystretch}{0.8}\begin{tabular}{@{}c@{}}1.230\\\mbox{} \end{tabular}&\renewcommand{\arraystretch}{0.8}\begin{tabular}{@{}c@{}}\mbox{}\\\end{tabular}&\renewcommand{\arraystretch}{0.8}\begin{tabular}{@{}c@{}}-0.000(17)\\0.000(17)\end{tabular}\\
\end{longtable}
\hypertarget{foot1}{\textsuperscript{1}} Inconsistent data as listed in Table.~\ref{tab:wrongdata}.

\hypertarget{foot2}{\textsuperscript{2}} Data with insufficient $q$ covered in experiments as listed in Table.~\ref{tab: outliers}.

\hypertarget{foot3}{\textsuperscript{3}} No uncertainties were given in original references.

\newpage
\textbf{Table IV. Fourier-Bessel coefficients. \\\\}
\begin{tabular}{@{}p{0.0in}p{1.5in}p{4.5in}@{}}
& $^{1}\langle r^2 \rangle^{1/2}_p$ & rms point-proton radius calculated from $\langle r^2 \rangle^{1/2}_c$ \\
& $^{2}\langle r^2 \rangle^{1/2}_p$ & rms point-proton radius calculated from the deduced $\rho_{p}$\\
& $\Delta \langle r^2 \rangle_{p}^{1/2}$ &  $^{2}\langle r^2 \rangle^{1/2}_p$-$^{1}\langle r^2 \rangle^{1/2}_p$\\
& a1...a17 & List of the Fourier-Bessel coefficients $a_{\nu}$, with $\nu$ = 1 to 17\\
& $R_{\mathrm{cut}}$ & The cutoff radius of $\rho_p$, beyond which the point-proton density is assumed to be identical to zero\\
\\
\end{tabular}

\renewcommand{\arraystretch}{1.3}

\begin{table}[H] 
\begin{ruledtabular}
\caption{Fourier-Bessel Coefficients. Data marked with * means the original $\rho_c$ is SOG.\label{tab:grossdatafb}} 
\begin{tabular}{ccccccc} 
\parbox{2.2cm}{\centering Nucleus \\ $^{1}\langle r^2 \rangle_{p}^{1/2}$ (fm) $^{2}\langle r^2 \rangle_{p}^{1/2}$ (fm) \ $\Delta  \langle r^2 \rangle_{p}^{1/2}$ (fm)} & 
\parbox{2.2cm}{\centering$^{12}\mathrm{C}$\\2.348(15) \\2.346(15)\\-0.002(22)}& 
\parbox{2.2cm}{\centering$^{12}\mathrm{C}$\\2.339(12) \\2.339(12)\\0.000(17)}& 
\parbox{2.2cm}{\centering$^{12}\mathrm{C}$\\2.354(5) \\2.353(5)\\-0.001(8)}& 
\parbox{2.2cm}{\centering$^{12}\mathrm{C}^*$\\2.345(6) \\2.346(6)\\0.001(8)}& 
\parbox{2.2cm}{\centering$^{15}\mathrm{N}$\\2.497(9) \\2.496(9)\\-0.001(13)}& 
\parbox{2.2cm}{\centering$^{15}\mathrm{N}$\\2.499(9) \\2.499(9)\\0.000(13)}\\ 
\colrule 
a1&1.596563e-02&1.598004e-02&1.595297e-02&2.281975e-02&2.599638e-02&2.599241e-02\\ 
a2&4.115996e-02&4.133531e-02&4.103032e-02&5.136918e-02&5.468020e-02&5.469316e-02\\ 
a3&4.212352e-02&4.244053e-02&4.167720e-02&3.926107e-02&3.537719e-02&3.539854e-02\\ 
a4&1.855897e-02&1.871583e-02&1.808079e-02&3.571197e-03&-7.424451e-03&-7.438440e-03\\ 
a5&-6.190735e-03&-6.413033e-03&-6.383995e-03&-1.771289e-02&-2.493136e-02&-2.491100e-02\\ 
a6&-1.611385e-02&-1.657838e-02&-1.622396e-02&-1.637021e-02&-1.419855e-02&-1.435063e-02\\ 
a7&-1.328560e-02&-1.324473e-02&-1.282481e-02&-7.016646e-03&-5.401176e-03&-5.177853e-03\\ 
a8&-6.299771e-03&-6.569263e-03&-6.170953e-03&-1.698104e-03&-1.316962e-03&-1.098842e-03\\ 
a9&-1.726897e-03&-2.098014e-03&-1.540239e-03&-3.723297e-04& & \\ 
a10&2.335173e-04&2.175797e-04&-8.927504e-04&7.794712e-04& & \\ 
a11&7.194713e-04&4.149610e-04&-6.668794e-04&1.570835e-03& & \\ 
a12&3.514963e-04&-1.726156e-04&5.821748e-04&1.151321e-03& & \\ 
a13&2.866935e-04&-2.814502e-05&-2.722491e-04& & & \\ 
a14&9.609303e-05&6.350011e-05&1.052738e-04& & & \\ 
a15&1.856198e-05&-3.820307e-05&-3.620588e-05& & & \\ 
a16&6.520046e-06& &1.135163e-05& & & \\ 
a17& & &-3.277771e-06& & & \\ 
$R_{\mathrm{cut}}$ (fm)&8.0&8.0&8.0&7.0&7.0&7.0\\ 
\end{tabular} 
\end{ruledtabular} 
\end{table} 
\begin{table}[H] 
\begin{ruledtabular}
\begin{tabular}{ccccccc} 
\parbox{2.2cm}{\centering Nucleus \\ $^{1}\langle r^2 \rangle_{p}^{1/2}$ (fm) $^{2}\langle r^2 \rangle_{p}^{1/2}$ (fm) \ $\Delta  \langle r^2 \rangle_{p}^{1/2}$ (fm)} & 
\parbox{2.2cm}{\centering$^{16}\mathrm{O}$\hyperlink{foot1}{\textsuperscript{1}}\\2.625(8) \\2.628(8)\\0.003(11)}& 
\parbox{2.2cm}{\centering$^{16}\mathrm{O}^*$\\2.598 \\2.597\\-0.001}& 
\parbox{2.2cm}{\centering$^{24}\mathrm{Mg}^*$\\2.927 \\2.927\\0.000}& 
\parbox{2.2cm}{\centering$^{26}\mathrm{Mg}$\\2.862(4) \\2.862(4)\\0.000(5)}& 
\parbox{2.2cm}{\centering$^{27}\mathrm{Al}$\\2.937(2) \\2.936(2)\\-0.001(2)}& 
\parbox{2.2cm}{\centering$^{28}\mathrm{Si}$\\2.986(17) \\2.986(17)\\0.000(24)}\\ 
\colrule 
a1&2.054637e-02&2.914617e-02&3.463227e-02&2.971836e-02&4.428937e-02&3.401273e-02\\ 
a2&4.760091e-02&5.647782e-02&6.037433e-02&5.889596e-02&6.519556e-02&6.326490e-02\\ 
a3&3.837557e-02&2.778930e-02&1.924068e-02&2.726175e-02&3.440602e-03&2.400862e-02\\ 
a4&4.431331e-03&-1.378297e-02&-1.991819e-02&-1.752891e-02&-3.173175e-02&-2.137867e-02\\ 
a5&-1.758502e-02&-2.337924e-02&-1.639629e-02&-2.246972e-02&-1.255088e-02&-2.145450e-02\\ 
a6&-1.702676e-02&-1.069532e-02&-1.411800e-03&-5.818969e-03&4.309743e-03&-1.537139e-03\\ 
a7&-6.562210e-03&9.608988e-04&2.623029e-03&3.743963e-03&2.444201e-03&6.413753e-03\\ 
a8&-9.490810e-04&3.272896e-03&1.826016e-03&3.250719e-03&3.477012e-04&1.350742e-03\\ 
a9&-2.565435e-03&-4.063177e-06&1.448918e-03&-2.320072e-04&-4.422203e-04&-1.086999e-03\\ 
a10&-8.380531e-04&-2.066841e-03&2.718820e-04&-4.471895e-04&2.364777e-04&3.981004e-04\\ 
a11&9.269518e-04&-1.211414e-03&-1.451154e-03&2.679222e-04&-9.267005e-05&-5.069579e-05\\ 
a12&-4.964851e-04&-9.952122e-05&-1.948824e-03&-9.177517e-05&2.708646e-05&-4.344608e-05\\ 
a13&2.328624e-04&1.326208e-04&-1.063502e-03&1.648929e-05& &4.075372e-05\\ 
a14& & & &6.570403e-07& &-2.827182e-05\\ 
a15& & & &-1.566440e-06& & \\ 
a16& & & & & & \\ 
a17& & & & & & \\ 
$R_{\mathrm{cut}}$ (fm)&8.0&7.0&7.5&8.0&7.0&8.0\\ 
\end{tabular} 
\end{ruledtabular} 
\end{table} 
\begin{table}[H] 
\begin{ruledtabular}
\begin{tabular}{ccccccc} 
\parbox{2.2cm}{\centering Nucleus \\ $^{1}\langle r^2 \rangle_{p}^{1/2}$ (fm) $^{2}\langle r^2 \rangle_{p}^{1/2}$ (fm) \ $\Delta  \langle r^2 \rangle_{p}^{1/2}$ (fm)} & 
\parbox{2.2cm}{\centering$^{28}\mathrm{Si}^*$\\3.056 \\3.057\\0.001}& 
\parbox{2.2cm}{\centering$^{29}\mathrm{Si}$\\2.983(17) \\2.983(17)\\0.000(24)}& 
\parbox{2.2cm}{\centering$^{30}\mathrm{Si}$\\3.080(25) \\3.080(25)\\0.000(36)}& 
\parbox{2.2cm}{\centering$^{30}\mathrm{Si}$\\3.101(13) \\3.099(13)\\-0.002(18)}& 
\parbox{2.2cm}{\centering$^{31}\mathrm{P}$\\3.093(10) \\3.092(10)\\-0.001(14)}& 
\parbox{2.2cm}{\centering$^{31}\mathrm{P}$\\3.097(5) \\3.096(5)\\-0.001(7)}\\ 
\colrule 
a1&3.365691e-02&3.403109e-02&2.877670e-02&2.870428e-02&3.584724e-02&3.582275e-02\\ 
a2&6.100640e-02&6.337192e-02&5.708663e-02&5.662182e-02&6.333583e-02&6.323283e-02\\ 
a3&2.185060e-02&2.352947e-02&2.829021e-02&2.772141e-02&1.973928e-02&1.970157e-02\\ 
a4&-2.112481e-02&-2.352680e-02&-1.577822e-02&-1.581504e-02&-2.435994e-02&-2.440411e-02\\ 
a5&-2.007345e-02&-2.358947e-02&-2.406060e-02&-2.340410e-02&-1.931094e-02&-1.877773e-02\\ 
a6&-2.052967e-03&-1.956243e-03&-7.250047e-03&-6.863818e-03&1.037494e-03&2.308588e-03\\ 
a7&5.133502e-03&5.373473e-03&4.430734e-03&5.779925e-03&6.801934e-03&7.048409e-03\\ 
a8&2.342732e-03&-1.525890e-04&3.069529e-03&4.615731e-03&1.895970e-03&1.448863e-03\\ 
a9&-4.902053e-04&-9.338247e-04&-7.364589e-04&9.983989e-04&-1.292787e-03&-8.716005e-04\\ 
a10&-1.700314e-03&5.682957e-04&1.393181e-04&5.867981e-04&6.166283e-04&5.903799e-04\\ 
a11&-2.367768e-03&-2.212257e-05&0.000000e+00&-8.912147e-04&-2.695010e-04&-4.173896e-04\\ 
a12&-1.717107e-03&5.742443e-05&-2.087785e-05&9.353538e-04&1.077349e-04&3.094568e-04\\ 
a13&-5.268889e-04& &1.453235e-05&-9.054803e-04&-4.329455e-05&-2.393001e-04\\ 
a14&3.191184e-05& & & & &1.917220e-04\\ 
a15& & & & & &-1.582329e-04\\ 
a16& & & & & & \\ 
a17& & & & & & \\ 
$R_{\mathrm{cut}}$ (fm)&8.0&8.0&8.5&8.5&8.0&8.0\\ 
\end{tabular} 
\end{ruledtabular} 
\end{table} 
\begin{table}[H] 
\begin{ruledtabular}
\begin{tabular}{ccccccc} 
\parbox{2.2cm}{\centering Nucleus \\ $^{1}\langle r^2 \rangle_{p}^{1/2}$ (fm) $^{2}\langle r^2 \rangle_{p}^{1/2}$ (fm) \ $\Delta  \langle r^2 \rangle_{p}^{1/2}$ (fm)} & 
\parbox{2.2cm}{\centering$^{32}\mathrm{S}$\\3.155(4) \\3.155(4)\\0.000(5)}& 
\parbox{2.2cm}{\centering$^{32}\mathrm{S}$\\3.134(5) \\3.132(5)\\-0.002(7)}& 
\parbox{2.2cm}{\centering$^{32}\mathrm{S}^*$\\3.165 \\3.163\\-0.002}& 
\parbox{2.2cm}{\centering$^{34}\mathrm{S}$\\3.190(4) \\3.190(4)\\0.000(5)}& 
\parbox{2.2cm}{\centering$^{36}\mathrm{S}$\\3.190(6) \\3.189(6)\\-0.001(8)}& 
\parbox{2.2cm}{\centering$^{40}\mathrm{Ar}$\\3.339(14) \\3.339(14)\\0.000(20)}\\ 
\colrule 
a1&3.782632e-02&3.794495e-02&3.249026e-02&3.759484e-02&3.757796e-02&3.080704e-02\\ 
a2&6.402472e-02&6.473672e-02&6.177866e-02&6.209099e-02&6.140773e-02&5.795718e-02\\ 
a3&1.687779e-02&1.719937e-02&2.653505e-02&1.378933e-02&1.143655e-02&2.239808e-02\\ 
a4&-2.321547e-02&-2.372906e-02&-1.794419e-02&-2.395507e-02&-2.488865e-02&-2.009017e-02\\ 
a5&-1.480129e-02&-1.455381e-02&-1.914384e-02&-1.185839e-02&-9.489573e-03&-1.792485e-02\\ 
a6&5.021319e-03&5.127808e-03&-4.303782e-04&7.429517e-03&1.003871e-02&-6.388636e-05\\ 
a7&6.801478e-03&7.559575e-03&7.909112e-03&5.412630e-03&7.148188e-03&1.545077e-03\\
\end{tabular} 
\end{ruledtabular} 
\end{table} 
\begin{table}[H] 
\begin{ruledtabular}
\begin{tabular}{ccccccc} 
\parbox{2.2cm}{\centering Nucleus \\ $^{1}\langle r^2 \rangle_{p}^{1/2}$ (fm) $^{2}\langle r^2 \rangle_{p}^{1/2}$ (fm) \ $\Delta  \langle r^2 \rangle_{p}^{1/2}$ (fm)} & 
\parbox{2.2cm}{\centering$^{39}\mathrm{K}^*$\\3.340 \\3.341\\0.001}& 
\parbox{2.2cm}{\centering$^{40}\mathrm{Ca}$\\3.362(10) \\3.362(10)\\0.000(14)}& 
\parbox{2.2cm}{\centering$^{40}\mathrm{Ca}^*$\\3.393(3) \\3.394(3)\\0.001(4)}& 
\parbox{2.2cm}{\centering$^{48}\mathrm{Ca}$\\3.370(9) \\3.370(9)\\0.000(13)}& 
\parbox{2.2cm}{\centering$^{48}\mathrm{Ca}^*$\\3.379 \\3.380\\0.001}& 
\parbox{2.2cm}{\centering$^{48}\mathrm{Ti}$\\3.516(1) \\3.515(1)\\-0.001(1)}\\ 
\colrule 
a1&4.347822e-02&4.554013e-02&3.397428e-02&4.542292e-02&4.536497e-02&2.811708e-02\\ 
a2&6.460674e-02&6.517030e-02&6.241962e-02&6.298636e-02&6.315868e-02&5.756767e-02\\ 
a3&2.881319e-03&-1.924672e-03&2.123111e-02&-8.410696e-03&-7.660866e-03&2.869154e-02\\ 
a4&-3.156528e-02&-3.316478e-02&-2.564650e-02&-3.674488e-02&-3.745994e-02&-1.982559e-02\\ 
a5&-5.873454e-03&-4.252134e-03&-2.054689e-02&-3.975087e-04&-9.202509e-04&-2.739512e-02\\ 
a6&1.336223e-02&1.409978e-02&4.848680e-03&1.740118e-02&1.766053e-02&-3.438404e-03\\ 
a7&7.677105e-03&6.810154e-03&1.217437e-02&5.696771e-03&5.735914e-03&1.178502e-02\\
\end{tabular} 
\end{ruledtabular} 
\end{table} 
\begin{table}[H] 
\begin{ruledtabular}
\begin{tabular}{ccccccc} 
\parbox{2.2cm}{\centering Nucleus \\ $^{1}\langle r^2 \rangle_{p}^{1/2}$ (fm) $^{2}\langle r^2 \rangle_{p}^{1/2}$ (fm) \ $\Delta  \langle r^2 \rangle_{p}^{1/2}$ (fm)} & 
\parbox{2.2cm}{\centering$^{50}\mathrm{Ti}$\\3.492(2) \\3.490(2)\\-0.002(2)}& 
\parbox{2.2cm}{\centering$^{50}\mathrm{Cr}$\hyperlink{foot1}{\textsuperscript{1}}\\3.581(4) \\3.587(4)\\0.006(5)}& 
\parbox{2.2cm}{\centering$^{52}\mathrm{Cr}$\hyperlink{foot1}{\textsuperscript{1}}\\3.563(3) \\3.570(3)\\0.007(4)}& 
\parbox{2.2cm}{\centering$^{54}\mathrm{Cr}$\hyperlink{foot1}{\textsuperscript{1}}\\3.611(4) \\3.615(4)\\0.004(5)}& 
\parbox{2.2cm}{\centering$^{54}\mathrm{Fe}$\\3.582(25) \\3.583(25)\\0.001(36)}& 
\parbox{2.2cm}{\centering$^{56}\mathrm{Fe}$\\3.635(24) \\3.635(24)\\0.000(34)}\\ 
\colrule 
a1&3.215388e-02&3.962133e-02&3.972388e-02&3.944157e-02&4.284514e-02&4.251579e-02\\ 
a2&6.101673e-02&6.482571e-02&6.546614e-02&6.314552e-02&6.756200e-02&6.532656e-02\\ 
a3&2.152733e-02&7.618844e-03&6.934007e-03&5.080074e-03&1.760707e-03&2.663486e-04\\ 
a4&-2.857098e-02&-3.630297e-02&-3.947223e-02&-3.678638e-02&-4.232862e-02&-3.935634e-02\\ 
a5&-2.403903e-02&-1.311235e-02&-1.408978e-02&-1.204605e-02&-1.343669e-02&-1.058378e-02\\ 
a6&4.793602e-03&1.310386e-02&1.560344e-02&1.376826e-02&1.798943e-02&1.609531e-02\\ 
a7&1.435131e-02&1.161340e-02&1.441065e-02&1.015703e-02&1.055464e-02&8.288705e-03\\
\end{tabular} 
\end{ruledtabular} 
\end{table} 
\begin{table}[H] 
\begin{ruledtabular}
\begin{tabular}{ccccccc} 
\parbox{2.2cm}{\centering Nucleus \\ $^{1}\langle r^2 \rangle_{p}^{1/2}$ (fm) $^{2}\langle r^2 \rangle_{p}^{1/2}$ (fm) \ $\Delta  \langle r^2 \rangle_{p}^{1/2}$ (fm)} & 
\parbox{2.2cm}{\centering$^{58}\mathrm{Fe}$\\3.669(25) \\3.670(25)\\0.001(36)}& 
\parbox{2.2cm}{\centering$^{59}\mathrm{Co}$\\3.711(5) \\3.711(5)\\0.000(7)}& 
\parbox{2.2cm}{\centering$^{58}\mathrm{Ni}$\\3.690(13) \\3.689(13)\\-0.001(18)}& 
\parbox{2.2cm}{\centering$^{58}\mathrm{Ni}$\\3.662(24) \\3.663(24)\\0.001(34)}& 
\parbox{2.2cm}{\centering$^{58}\mathrm{Ni}^*$\\3.693(4) \\3.695(4)\\0.002(5)}& 
\parbox{2.2cm}{\centering$^{60}\mathrm{Ni}$\\3.719(13) \\3.719(13)\\0.000(18)}\\ 
\colrule 
a1&4.227483e-02&4.364114e-02&4.542231e-02&4.557003e-02&5.195915e-02&4.520016e-02\\ 
a2&6.338518e-02&6.416929e-02&6.790913e-02&6.821116e-02&6.586682e-02&6.610298e-02\\ 
a3&-1.660283e-03&-3.608071e-03&-3.101441e-03&-3.651049e-03&-1.802876e-02&-4.751480e-03\\ 
a4&-3.735839e-02&-3.920587e-02&-4.455637e-02&-4.362350e-02&-4.132518e-02&-4.180750e-02\\ 
a5&-7.655770e-03&-6.555865e-03&-9.554412e-03&-8.960143e-03&4.071859e-03&-6.440688e-03\\ 
a6&1.568331e-02&1.703081e-02&2.026865e-02&1.960069e-02&2.023399e-02&1.900107e-02\\ 
a7&6.988238e-03&9.332996e-03&1.125435e-02&8.636691e-03&4.359107e-03&9.186037e-03\\
\end{tabular} 
\end{ruledtabular} 
\end{table} 
\begin{table}[H] 
\begin{ruledtabular}
\begin{tabular}{ccccccc} 
\parbox{2.2cm}{\centering Nucleus \\ $^{1}\langle r^2 \rangle_{p}^{1/2}$ (fm) $^{2}\langle r^2 \rangle_{p}^{1/2}$ (fm) \ $\Delta  \langle r^2 \rangle_{p}^{1/2}$ (fm)} & 
\parbox{2.2cm}{\centering$^{60}\mathrm{Ni}$\\3.686(24) \\3.687(24)\\0.001(34)}& 
\parbox{2.2cm}{\centering$^{60}\mathrm{Ni}$\\3.704(24) \\3.705(24)\\0.001(34)}& 
\parbox{2.2cm}{\centering$^{62}\mathrm{Ni}$\\3.726(24) \\3.726(24)\\0.000(34)}& 
\parbox{2.2cm}{\centering$^{64}\mathrm{Ni}$\hyperlink{foot1}{\textsuperscript{1}}\\3.747(24) \\3.741(24)\\-0.006(34)}& 
\parbox{2.2cm}{\centering$^{63}\mathrm{Cu}$\\3.810(5) \\3.809(5)\\-0.001(7)}& 
\parbox{2.2cm}{\centering$^{65}\mathrm{Cu}$\\3.831(5) \\3.830(5)\\-0.001(7)}\\ 
\colrule 
a1&4.538472e-02&4.527013e-02&4.510057e-02&4.496973e-02&4.613848e-02&4.597753e-02\\ 
a2&6.652640e-02&6.601390e-02&6.439329e-02&6.292853e-02&6.360731e-02&6.235324e-02\\ 
a3&-5.662318e-03&-5.536085e-03&-7.697912e-03&-1.018829e-02&-8.723592e-03&-1.052501e-02\\ 
a4&-4.143514e-02&-4.120692e-02&-3.970552e-02&-4.001176e-02&-3.796898e-02&-3.779939e-02\\ 
a5&-5.713106e-03&-5.836025e-03&-3.296697e-03&-6.959134e-04&-1.911211e-03&3.020420e-04\\ 
a6&1.896097e-02&1.902484e-02&1.874935e-02&1.938670e-02&1.624056e-02&1.652803e-02\\ 
a7&8.431918e-03&7.921799e-03&6.242866e-03&5.879437e-03&7.175426e-03&6.267175e-03\\
\end{tabular} 
\end{ruledtabular} 
\end{table} 
\begin{table}[H] 
\begin{ruledtabular}
\begin{tabular}{ccccccc} 
\parbox{2.2cm}{\centering Nucleus \\ $^{1}\langle r^2 \rangle_{p}^{1/2}$ (fm) $^{2}\langle r^2 \rangle_{p}^{1/2}$ (fm) \ $\Delta  \langle r^2 \rangle_{p}^{1/2}$ (fm)} & 
\parbox{2.2cm}{\centering$^{64}\mathrm{Zn}$\\3.823(23) \\3.824(23)\\0.001(33)}& 
\parbox{2.2cm}{\centering$^{66}\mathrm{Zn}$\hyperlink{foot1}{\textsuperscript{1}}\\3.829(25) \\3.826(25)\\-0.003(36)}& 
\parbox{2.2cm}{\centering$^{68}\mathrm{Zn}$\\3.875(24) \\3.875(24)\\0.000(34)}& 
\parbox{2.2cm}{\centering$^{70}\mathrm{Zn}$\\3.916(31) \\3.915(31)\\-0.001(44)}& 
\parbox{2.2cm}{\centering$^{70}\mathrm{Ge}$\\3.971(2) \\3.971(2)\\0.000(2)}& 
\parbox{2.2cm}{\centering$^{72}\mathrm{Ge}$\\3.989(2) \\3.989(2)\\0.000(2)}\\ 
\colrule 
a1&4.759588e-02&4.755670e-02&4.719818e-02&4.689987e-02&3.854705e-02&3.844216e-02\\ 
a2&6.449839e-02&6.389514e-02&6.158918e-02&5.977867e-02&6.262698e-02&6.159914e-02\\ 
a3&-9.998843e-03&-1.072417e-02&-1.360669e-02&-1.535406e-02&7.000757e-03&5.186830e-03\\ 
a4&-3.684871e-02&-3.652183e-02&-3.574146e-02&-3.586833e-02&-3.413056e-02&-3.468691e-02\\ 
a5&-1.042917e-03&-7.060767e-04&3.382339e-03&4.643790e-03&-1.204848e-02&-1.109233e-02\\ 
a6&1.490984e-02&1.437325e-02&1.509797e-02&1.486399e-02&1.213304e-02&1.324801e-02\\ 
a7&2.374474e-03&2.370305e-03&5.335735e-03&6.840799e-03&7.561921e-03&7.337414e-03\\
\end{tabular} 
\end{ruledtabular} 
\end{table} 
\begin{table}[H] 
\begin{ruledtabular}
\begin{tabular}{ccccccc} 
\parbox{2.2cm}{\centering Nucleus \\ $^{1}\langle r^2 \rangle_{p}^{1/2}$ (fm) $^{2}\langle r^2 \rangle_{p}^{1/2}$ (fm) \ $\Delta  \langle r^2 \rangle_{p}^{1/2}$ (fm)} & 
\parbox{2.2cm}{\centering$^{74}\mathrm{Ge}$\\4.005(2) \\4.005(2)\\0.000(2)}& 
\parbox{2.2cm}{\centering$^{76}\mathrm{Ge}$\\4.012(2) \\4.012(2)\\0.000(2)}& 
\parbox{2.2cm}{\centering$^{88}\mathrm{Sr}$\\4.130(6) \\4.130(6)\\0.000(8)}& 
\parbox{2.2cm}{\centering$^{90}\mathrm{Zr}$\\4.191(8) \\4.191(8)\\0.000(11)}& 
\parbox{2.2cm}{\centering$^{92}\mathrm{Zr}$\\4.229(11) \\4.229(11)\\0.000(15)}& 
\parbox{2.2cm}{\centering$^{94}\mathrm{Zr}$\\4.250(10) \\4.250(10)\\0.000(14)}\\ 
\colrule 
a1&3.834233e-02&3.829970e-02&5.708150e-02&4.662302e-02&4.636760e-02&4.622108e-02\\ 
a2&6.048918e-02&6.002528e-02&5.763384e-02&6.414544e-02&6.236847e-02&6.145578e-02\\ 
a3&2.976200e-03&1.660314e-03&-3.692578e-02&-1.338610e-02&-1.449057e-02&-1.453191e-02\\ 
a4&-3.545776e-02&-3.673767e-02&-3.114379e-02&-4.274922e-02&-4.060408e-02&-3.841225e-02\\ 
a5&-1.019560e-02&-9.595041e-03&2.070933e-02&3.156215e-03&3.974682e-03&4.982994e-03\\ 
a6&1.404737e-02&1.538349e-02&1.106822e-02&2.097814e-02&1.889424e-02&1.747524e-02\\ 
a7&7.536844e-03&7.722211e-03&-9.188192e-03&-8.447194e-04&-1.262836e-03&-2.405147e-03\\
\end{tabular} 
\end{ruledtabular} 
\end{table} 
\begin{table}[H] 
\begin{ruledtabular}
\begin{tabular}{ccccccc} 
\parbox{2.2cm}{\centering Nucleus \\ $^{1}\langle r^2 \rangle_{p}^{1/2}$ (fm) $^{2}\langle r^2 \rangle_{p}^{1/2}$ (fm) \ $\Delta  \langle r^2 \rangle_{p}^{1/2}$ (fm)} & 
\parbox{2.2cm}{\centering$^{92}\mathrm{Mo}$\\4.227(16) \\4.227(16)\\0.000(22)}& 
\parbox{2.2cm}{\centering$^{94}\mathrm{Mo}$\\4.266(16) \\4.265(16)\\-0.001(22)}& 
\parbox{2.2cm}{\centering$^{96}\mathrm{Mo}$\\4.299(16) \\4.299(16)\\0.000(22)}& 
\parbox{2.2cm}{\centering$^{98}\mathrm{Mo}$\hyperlink{foot1}{\textsuperscript{1}}\\4.324(16) \\4.320(16)\\-0.004(22)}& 
\parbox{2.2cm}{\centering$^{100}\mathrm{Mo}$\\4.366(16) \\4.366(16)\\0.000(22)}& 
\parbox{2.2cm}{\centering$^{104}\mathrm{Pd}$\\4.373(10) \\4.373(10)\\0.000(14)}\\ 
\colrule 
a1&3.098536e-02&3.086412e-02&3.076239e-02&3.068550e-02&3.054627e-02&4.152979e-02\\ 
a2&6.148961e-02&6.037915e-02&5.952882e-02&5.868812e-02&5.752575e-02&6.481298e-02\\ 
a3&2.334881e-02&2.161981e-02&2.040004e-02&1.894219e-02&1.699471e-02&-2.271533e-03\\ 
a4&-3.210825e-02&-3.195330e-02&-3.141877e-02&-3.140961e-02&-3.180175e-02&-4.331657e-02\\ 
a5&-3.135557e-02&-2.940467e-02&-2.729121e-02&-2.547199e-02&-2.416081e-02&-5.392620e-03\\ 
a6&5.079051e-03&5.614998e-03&6.247289e-03&7.097433e-03&7.790335e-03&2.174538e-02\\ 
a7&1.959579e-02&1.778918e-02&1.639736e-02&1.516128e-02&1.488918e-02&5.267997e-03\\
\end{tabular} 
\end{ruledtabular} 
\end{table} 
\begin{table}[H] 
\begin{ruledtabular}
\begin{tabular}{ccccccc} 
\parbox{2.2cm}{\centering Nucleus \\ $^{1}\langle r^2 \rangle_{p}^{1/2}$ (fm) $^{2}\langle r^2 \rangle_{p}^{1/2}$ (fm) \ $\Delta  \langle r^2 \rangle_{p}^{1/2}$ (fm)} & 
\parbox{2.2cm}{\centering$^{106}\mathrm{Pd}$\\4.404(11) \\4.404(11)\\0.000(15)}& 
\parbox{2.2cm}{\centering$^{108}\mathrm{Pd}$\\4.462(10) \\4.462(10)\\0.000(14)}& 
\parbox{2.2cm}{\centering$^{110}\mathrm{Pd}$\\4.479(10) \\4.479(10)\\0.000(14)}& 
\parbox{2.2cm}{\centering$^{116}\mathrm{Sn}^*$\\4.565(1) \\4.565(1)\\0.000(1)}& 
\parbox{2.2cm}{\centering$^{124}\mathrm{Sn}^*$\\4.618(1) \\4.619(1)\\0.001(1)}& 
\parbox{2.2cm}{\centering$^{144}\mathrm{Sm}$\hyperlink{foot1}{\textsuperscript{1}}\\4.886(9) \\4.891(9)\\0.005(12)}\\ 
\colrule 
a1&4.137188e-02&4.106493e-02&4.097561e-02&6.899015e-02&6.817262e-02&7.548876e-02\\ 
a2&6.367390e-02&6.129021e-02&6.058746e-02&4.043861e-02&3.541013e-02&2.729386e-02\\ 
a3&-3.200717e-03&-5.787408e-03&-6.564896e-03&-5.747236e-02&-5.918981e-02&-7.026464e-02\\ 
a4&-4.214550e-02&-4.092247e-02&-4.052263e-02&-2.648813e-03&2.235317e-03&1.234955e-02\\ 
a5&-4.259386e-03&-2.645883e-03&-2.096760e-03&3.001300e-02&2.901410e-02&2.486867e-02\\ 
a6&2.096266e-02&2.005782e-02&1.947694e-02&-9.331048e-03&-1.252691e-02&-1.813711e-02\\ 
a7&4.063007e-03&3.687753e-03&2.837584e-03&-1.112177e-02&-9.868878e-03&-6.462613e-03\\ 
a8&-8.781408e-03&-8.289686e-03&-8.328761e-03&2.799850e-03&3.780635e-03&-3.472963e-03\\ 
a9&-2.707365e-03&-3.455343e-03&-2.491436e-03&4.358680e-04&3.398852e-04&2.693322e-03\\ 
a10&4.384550e-03&2.027586e-03&3.133509e-03&-1.073731e-03&-8.524031e-04&-2.946864e-03\\ 
a11&2.921031e-03&5.061815e-04&6.955663e-04&6.307835e-04&7.014318e-04&-4.524552e-03\\ 
a12&4.150467e-05&-8.379280e-04&-1.052261e-03&1.780342e-04&1.380447e-04& \\ 
a13& & & &-2.351546e-04&-2.454906e-04& \\ 
a14& & & &-1.550445e-05&-2.839644e-05& \\ 
a15& & & &4.106462e-05&3.271429e-05& \\ 
a16& & & & & & \\ 
a17& & & & & & \\ 
$R_{\mathrm{cut}}$ (fm)&11.0&11.0&11.0&9.0&9.0&9.25\\ 
\end{tabular} 
\end{ruledtabular} 
\end{table} 
\begin{table}[H] 
\begin{ruledtabular}
\begin{tabular}{ccccccc} 
\parbox{2.2cm}{\centering Nucleus \\ $^{1}\langle r^2 \rangle_{p}^{1/2}$ (fm) $^{2}\langle r^2 \rangle_{p}^{1/2}$ (fm) \ $\Delta  \langle r^2 \rangle_{p}^{1/2}$ (fm)} & 
\parbox{2.2cm}{\centering$^{148}\mathrm{Sm}$\\4.921(8) \\4.921(8)\\0.000(11)}& 
\parbox{2.2cm}{\centering$^{148}\mathrm{Sm}$\hyperlink{foot1}{\textsuperscript{1}}\\4.947(6) \\4.952(6)\\0.005(8)}& 
\parbox{2.2cm}{\centering$^{150}\mathrm{Sm}$\hyperlink{foot1}{\textsuperscript{1}}\\4.995(6) \\5.008(6)\\0.013(8)}& 
\parbox{2.2cm}{\centering$^{152}\mathrm{Sm}$\\5.046(8) \\5.047(8)\\0.001(11)}& 
\parbox{2.2cm}{\centering$^{152}\mathrm{Sm}$\hyperlink{foot1}{\textsuperscript{1}}\\5.033(6) \\5.050(6)\\0.017(8)}& 
\parbox{2.2cm}{\centering$^{154}\mathrm{Sm}$\\5.073(8) \\5.074(8)\\0.001(11)}\\ 
\colrule 
a1&7.121389e-02&7.459119e-02&7.397074e-02&5.654716e-02&7.320934e-02&5.630702e-02\\ 
a2&3.394128e-02&2.506559e-02&2.568809e-02&4.661842e-02&2.275933e-02&4.545102e-02\\ 
a3&-6.064289e-02&-6.535464e-02&-5.799521e-02&-4.336271e-02&-5.943398e-02&-4.337816e-02\\ 
a4&5.871558e-03&1.271592e-02&1.249307e-02&-2.066911e-02&1.159723e-02&-2.044715e-02\\ 
a5&2.580718e-02&2.201007e-02&1.982634e-02&2.503678e-02&2.090990e-02&2.415934e-02\\ 
a6&-1.213987e-02&-1.639238e-02&-1.372574e-02&6.505643e-03&-9.402343e-03&6.843739e-03\\ 
a7&-6.352906e-03&-2.934120e-03&3.041652e-03&-9.883164e-03&8.654119e-03&-8.778126e-03\\ 
a8&3.467213e-03&1.718829e-03&1.456761e-03&-3.073585e-03&-2.593756e-03&-3.747890e-03\\ 
a9&-1.021071e-03&2.083367e-03&2.141497e-03&2.792917e-03&-2.093478e-03&8.545233e-04\\ 
a10&-4.033010e-04&-3.101280e-03&-6.528588e-03& &-5.628200e-03& \\ 
a11& &-4.981671e-03&-5.223002e-03& & & \\ 
a12& & & & & & \\ 
a13& & & & & & \\ 
a14& & & & & & \\ 
a15& & & & & & \\ 
a16& & & & & & \\ 
a17& & & & & & \\ 
$R_{\mathrm{cut}}$ (fm)&9.5&9.25&9.25&10.5&9.25&10.5\\ 
\end{tabular} 
\end{ruledtabular} 
\end{table} 
\begin{table}[H] 
\begin{ruledtabular}
\begin{tabular}{ccccccc} 
\parbox{2.2cm}{\centering Nucleus \\ $^{1}\langle r^2 \rangle_{p}^{1/2}$ (fm) $^{2}\langle r^2 \rangle_{p}^{1/2}$ (fm) \ $\Delta  \langle r^2 \rangle_{p}^{1/2}$ (fm)} & 
\parbox{2.2cm}{\centering$^{154}\mathrm{Gd}$\\5.070 \\5.070\\0.000}& 
\parbox{2.2cm}{\centering$^{158}\mathrm{Gd}$\\5.119(6) \\5.121(6)\\0.002(8)}& 
\parbox{2.2cm}{\centering$^{166}\mathrm{Er}$\hyperlink{foot1}{\textsuperscript{1}}\\5.174(20) \\5.171(20)\\-0.003(28)}& 
\parbox{2.2cm}{\centering$^{174}\mathrm{Yb}$\hyperlink{foot1}{\textsuperscript{1}}\\5.365(30) \\5.368(30)\\0.003(42)}& 
\parbox{2.2cm}{\centering$^{175}\mathrm{Lu}$\\5.315(30) \\5.313(30)\\-0.002(42)}& 
\parbox{2.2cm}{\centering$^{192}\mathrm{Os}$\\5.362(4) \\5.361(4)\\-0.001(5)}\\ 
\colrule 
a1&6.442225e-02&5.767503e-02&5.485308e-02&5.482275e-02&5.603736e-02&5.948057e-02\\ 
a2&3.834809e-02&4.448327e-02&4.859056e-02&4.123343e-02&4.351386e-02&4.273128e-02\\ 
a3&-5.226323e-02&-4.516590e-02&-4.132008e-02&-4.872458e-02&-4.811958e-02&-5.328530e-02\\ 
a4&-5.888588e-03&-1.956044e-02&-2.213074e-02&-2.352633e-02&-2.191630e-02&-1.143540e-02\\ 
a5&2.469187e-02&2.431282e-02&2.650956e-02&2.451196e-02&2.699652e-02&3.478904e-02\\ 
a6&-1.128790e-03&6.770923e-03&1.020063e-02&3.500804e-03&5.047266e-03&-7.411160e-03\\ 
a7&-7.125687e-03&-1.074815e-02&-7.503547e-03&-8.543717e-03&-1.023620e-02&-1.301689e-02\\ 
a8& &-3.394482e-03&7.809361e-04&-2.479263e-04&4.962607e-04&5.619795e-03\\ 
a9& &3.980999e-03&9.036106e-04&2.067865e-03&-4.140015e-04&4.963567e-03\\ 
a10& & &-6.814356e-04& & &4.678924e-04\\ 
a11& & &2.632456e-04& & &-3.650963e-03\\ 
a12& & & & & &1.812840e-03\\ 
a13& & & & & & \\ 
a14& & & & & & \\ 
a15& & & & & & \\ 
a16& & & & & & \\ 
a17& & & & & & \\ 
$R_{\mathrm{cut}}$ (fm)&10.0&10.5&11.0&11.0&11.0&11.0\\ 
\end{tabular} 
\end{ruledtabular} 
\end{table} 
\begin{table}[H] 
\begin{ruledtabular}
\begin{tabular}{ccccccc} 
\parbox{2.2cm}{\centering Nucleus \\ $^{1}\langle r^2 \rangle_{p}^{1/2}$ (fm) $^{2}\langle r^2 \rangle_{p}^{1/2}$ (fm) \ $\Delta  \langle r^2 \rangle_{p}^{1/2}$ (fm)} & 
\parbox{2.2cm}{\centering$^{196}\mathrm{Pt}$\\5.333(20) \\5.333(20)\\0.000(28)}& 
\parbox{2.2cm}{\centering$^{204}\mathrm{Hg}$\\5.426(1) \\5.428(1)\\0.002(2)}& 
\parbox{2.2cm}{\centering$^{203}\mathrm{Ti}$\\5.414(5) \\5.414(5)\\0.000(7)}& 
\parbox{2.2cm}{\centering$^{205}\mathrm{Ti}$\\5.421(5) \\5.421(5)\\0.000(7)}& 
\parbox{2.2cm}{\centering$^{205}\mathrm{Ti}^*$\\5.430 \\5.429\\-0.001}& 
\parbox{2.2cm}{\centering$^{204}\mathrm{Pb}$\\5.430(2) \\5.430(2)\\0.000(2)}\\ 
\colrule 
a1&5.052942e-02&5.118442e-02&5.188767e-02&5.183645e-02&7.577868e-02&5.242745e-02\\ 
a2&5.506466e-02&5.193619e-02&5.285239e-02&5.243914e-02&1.780112e-02&5.308675e-02\\ 
a3&-3.700815e-02&-4.201835e-02&-4.153893e-02&-4.180283e-02&-6.464683e-02&-4.142965e-02\\ 
a4&-3.814659e-02&-3.461087e-02&-3.400158e-02&-3.320597e-02&2.957895e-02&-3.226501e-02\\ 
a5&2.743065e-02&3.270222e-02&3.200764e-02&3.211366e-02&1.564746e-02&3.377158e-02\\ 
a6&1.781094e-02&1.312358e-02&1.341104e-02&1.292393e-02&-2.645869e-02&1.372434e-02\\ 
a7&-1.776232e-02&-2.191821e-02&-2.131844e-02&-2.103147e-02&6.101018e-03&-1.955237e-02\\
\end{tabular} 
\end{ruledtabular} 
\end{table} 
\begin{table}[H] 
\begin{ruledtabular}
\begin{tabular}{ccccccc} 
\parbox{2.2cm}{\centering Nucleus \\ $^{1}\langle r^2 \rangle_{p}^{1/2}$ (fm) $^{2}\langle r^2 \rangle_{p}^{1/2}$ (fm) \ $\Delta  \langle r^2 \rangle_{p}^{1/2}$ (fm)} & 
\parbox{2.2cm}{\centering$^{206}\mathrm{Pb}$\\5.441(2) \\5.441(2)\\0.000(2)}& 
\parbox{2.2cm}{\centering$^{206}\mathrm{Pb}^*$\\5.440 \\5.440\\0.000}& 
\parbox{2.2cm}{\centering$^{207}\mathrm{Pb}$\\5.448(2) \\5.447(2)\\-0.001(2)}& 
\parbox{2.2cm}{\centering$^{208}\mathrm{Pb}$\\5.454(2) \\5.454(2)\\0.000(2)}& 
\parbox{2.2cm}{\centering$^{208}\mathrm{Pb}$\\5.450(1) \\5.449(1)\\-0.001(1)}& 
\parbox{2.2cm}{\centering$^{208}\mathrm{Pb}^*$\\5.454(2) \\5.453(2)\\-0.001(2)}\\ 
\colrule 
a1&5.234297e-02&7.654807e-02&5.230327e-02&5.225736e-02&6.319962e-02&7.635409e-02\\ 
a2&5.246954e-02&1.761205e-02&5.233215e-02&5.203080e-02&3.968508e-02&1.688965e-02\\ 
a3&-4.170539e-02&-6.362697e-02&-4.168733e-02&-4.189223e-02&-5.883590e-02&-6.362851e-02\\ 
a4&-3.132780e-02&3.185113e-02&-3.134633e-02&-3.110798e-02&-3.030298e-03&3.176930e-02\\ 
a5&3.359855e-02&1.623619e-02&3.373440e-02&3.363981e-02&3.711014e-02&1.554303e-02\\ 
a6&1.327156e-02&-2.402946e-02&1.282384e-02&1.227475e-02&-1.284376e-02&-2.534708e-02\\ 
a7&-1.947619e-02&8.662215e-03&-1.876857e-02&-1.923838e-02&-1.310369e-02&8.218625e-03\\
\end{tabular} 
\end{ruledtabular} 
\end{table} 
\begin{table} 
\begin{ruledtabular}
\begin{tabular}{ccccccc} 
\parbox{2.2cm}{\centering Nucleus \\ $^{1}\langle r^2 \rangle_{p}^{1/2}$ (fm)\\ $^{2}\langle r^2 \rangle_{p}^{1/2}$ (fm) \ $\Delta  \langle r^2 \rangle_{p}^{1/2}$ (fm)} & 
\parbox{2.2cm}{\centering$^{209}\mathrm{Bi}$\\5.469(4) \\5.469(4)\\0.000(5)}& 
\parbox{2.2cm}{\mbox{} \\\mbox{} \\\mbox{} }& 
\parbox{2.2cm}{\mbox{} \\\mbox{} \\\mbox{} }& 
\parbox{2.2cm}{\mbox{} \\\mbox{} \\\mbox{} }& 
\parbox{2.2cm}{\mbox{} \\\mbox{} \\\mbox{} }& 
\parbox{2.2cm}{\mbox{} \\\mbox{} \\\mbox{} }\\ 
\colrule 
a1&5.277296e-02& \mbox{}&\mbox{} &\mbox{} &\mbox{} &\mbox{} \\ 
a2&5.165831e-02& \mbox{}&\mbox{} &\mbox{} &\mbox{} &\mbox{} \\ 
a3&-4.334622e-02& \mbox{}&\mbox{} &\mbox{} &\mbox{} &\mbox{} \\ 
a4&-3.079738e-02& \mbox{}&\mbox{} &\mbox{} &\mbox{} &\mbox{} \\ 
a5&3.443720e-02& \mbox{}&\mbox{} &\mbox{} &\mbox{} &\mbox{} \\ 
a6&1.217190e-02& \mbox{}&\mbox{} &\mbox{} &\mbox{} &\mbox{} \\ 
a7&-1.997960e-02& \mbox{}&\mbox{} &\mbox{} &\mbox{} &\mbox{} \\ 
a8&-4.651801e-04& \mbox{}&\mbox{} &\mbox{} &\mbox{} &\mbox{} \\ 
a9&1.232202e-02& \mbox{}&\mbox{} &\mbox{} &\mbox{} &\mbox{} \\ 
a10&1.012790e-03& \mbox{}&\mbox{} &\mbox{} &\mbox{} &\mbox{} \\ 
a11&-1.612628e-03& \mbox{}&\mbox{} &\mbox{} &\mbox{} &\mbox{} \\ 
a12&5.606470e-04& \mbox{}&\mbox{} &\mbox{} &\mbox{} &\mbox{} \\ 
a13&-2.693546e-04& \mbox{}&\mbox{} &\mbox{} &\mbox{} &\mbox{} \\ 
a14&1.421720e-04& \mbox{}&\mbox{} &\mbox{} &\mbox{} &\mbox{} \\ 
a15&-7.625756e-05& \mbox{}&\mbox{} &\mbox{} &\mbox{} &\mbox{} \\ 
a16&4.113084e-05& \mbox{}&\mbox{} &\mbox{} &\mbox{} &\mbox{} \\ 
a17&-2.209467e-05& \mbox{}&\mbox{} &\mbox{} &\mbox{} &\mbox{} \\ 
$R_{\mathrm{cut}}$ (fm)&12.0& \mbox{}&\mbox{} &\mbox{} &\mbox{} &\mbox{} \\ 
\end{tabular} 
\end{ruledtabular}

\hypertarget{foot1}{\textsuperscript{1}}\raggedright Data with large $\Delta \langle r^2 \rangle_{p}^{1/2}$ as listed in Tab.~\ref{tab: outliers}.
\end{table}

% \footnotetext[1]{Data with large $\Delta xxxxx$ as listed in Tab.~\ref{tab: outliers}.} 
% \footnotetext[2]{from SOG $\rho_c$.} 
% \end{table} 

\end{widetext}

\bibliographystyle{apsrev4-1}
\bibliography{apssamp}

\end{document}